\documentclass[aps,prl,reprint,groupedaddress,superscriptaddress,showpacs,floatfix]{revtex4-1}
\usepackage[dvipsnames,svgnames,x11names]{xcolor}
\usepackage{graphicx}
\usepackage{hyperref}
\usepackage{color}
\usepackage{amsmath}
\usepackage{upgreek}

\makeatletter

\newcommand{\Rmnum}[1]{\expandafter\@slowromancap\romannumeral #1@}
\makeatother

\begin{document}

\title{Optical trapping of ion Coulomb crystals}
\author{Julian Schmidt}
\author{Alexander Lambrecht}
\author{Pascal Weckesser}
\author{Markus Debatin}
\author{Leon Karpa}
\author{Tobias Schaetz}
\affiliation{Albert-Ludwigs-Universit\"{a}t Freiburg, Physikalisches Institut, Hermann-Herder-Stra{\ss}e 3, 79104 Freiburg, Germany}
\date{\today}
\begin{abstract}

The electronic and motional degrees of freedom of trapped ions can be controlled and coherently coupled on the level of individual quanta. 
Assembling complex quantum systems ion by ion while keeping this unique level of control remains a challenging task. 
For many applications, linear chains of ions in conventional traps are ideally suited to address this problem. 
However, driven motion due to the magnetic or radio-frequency electric trapping fields sometimes limits the performance in one dimension and severely affects the extension to higher dimensional systems.
Here, we report on the trapping of multiple Barium ions in a single-beam optical dipole trap without radio-frequency or additional magnetic fields. 
We study the persistence of order in ensembles of up to six ions within the optical trap, measure their temperature and conclude that the ions form a linear chain, commonly called a one-dimensional Coulomb crystal.  
As a proof-of-concept demonstration, we access the collective motion and perform spectrometry of the normal modes in the optical trap.
Our system provides a platform which is free of driven motion and combines advantages of optical trapping, such as state-dependent confinement and nano-scale potentials, with the desirable properties of crystals of trapped ions, such as long-range interactions featuring collective motion. 
Starting with small numbers of ions, it has been proposed that these properties would allow the experimental study of many-body physics and the onset of structural quantum phase transitions between one- and two-dimensional crystals.
\end{abstract}

\maketitle
\section{\Rmnum{1}. Introduction}
Coulomb crystals are an intriguing form of matter. 
On the one hand, it is believed that they make up the core of white dwarves and the surface of neutron stars \cite{Drewsen2015}. 
On the other hand, they provide a versatile solid-state-like platform for applications that require a magnified lattice structure. 
While a solid-state system with nanometer-spaced particles can hardly be observed and controlled with single-site resolution, distances of ions in Coulomb crystals are on the order of $a \approx 10\,\upmu\text{m}$.
The corresponding $10^{12}$ times lower densities allow to manipulate each constituent individually.
\\
The ions are typically trapped in Paul \cite{Paul1990} and Penning traps \cite{Dehmelt1990}, which combine electrostatic with radio-frequency (rf) fields or magnetic fields, respectively.  
These traps provide stiff confinement to counteract the repulsive interaction of the positively charged ions. 
Ensembles with temperatures of thousands of Kelvin can be trapped and form one-component plasmas (OCP). 
Cooling the OCP leads to a phase transition to a crystal when the ratio of the Coulomb energy for mean ion distances and the average kinetic energy, $\Gamma_{\text{Plasma}} = E_{\text{Coul}}/E_{\text{kin}}$, exceeds a critical value. Depending on the dimensionality of the system, Molecular Dynamics (MD) simulations predict that this phase transition occurs at different $\Gamma_{\text{Plasma}}$, e.g. $\Gamma_{\text{Plasma}}^{\text{1D}} = \frac{q^2}{k_B \cdot T \cdot  a} >1$ ($\Gamma_{\text{Plasma}}^{\text{2D}}>128$ and $\Gamma_{\text{Plasma}}^{\text{3D}}>174$) \cite{Caillol1982,Dubin1999}, where $q$ is the charge of the particles, $T$ their temperature and $k_B$ the Boltzmann constant. For typical experimental parameters in 1D, temperatures below a critical value $T_c$ on the order of $50\,\text{mK}$ are required to reach the phase transition. 
\\
A handful of ion species available for direct laser cooling can be prepared at Doppler temperatures $T_D << T_c$, and sufficiently well isolated from the environment by levitating them in ultra-high vacuum. Other elements and molecular ions can also be embedded into the crystalline structure thanks to sympathetic cooling via Coulomb interaction \cite{Drewsen2015}. 
\\
Extensive work has allowed to extend the coherent control and coupling of external (motional) and internal (electronic) degrees of freedom from single ions to (short) linear chains of ions \cite{Leibfried2003,Ballance2016,Chou2010,Porras2004,Friedenauer2008,Zhang2017}.
\\
However, ions perform rf-driven motion in Paul traps and driven cyclotron motion in Penning traps, which often is undesirable. Various methods can minimize driven motion close to zero \cite{Berkeland1998}, but, even when assuming perfect compensation of stray electric fields, driven motion remains inevitable if ions are intrinsically displaced from the trap center, e.g. in a 2D or 3D crystal \cite{Drewsen2015, Thompson2015}, or during the interaction with neutral atoms \cite{Cetina2012, Meir2016}. In these cases, the kinetic energy of the driven motion easily exceeds the residual thermal energy by orders of magnitude. This makes it challenging to extend the unique level of control and isolation available for single ions and linear chains of ions to Coulomb crystals of larger size and dimensionality. A generic approach to entirely mitigate driven motion is the use of optical dipole traps without rf and without magnetic fields. Optical dipole traps have been established in experiments for neutral particles for decades \cite{Grimm2000}.\\
Recently, trapping and isolation of a single ion in a dipole trap was demonstrated for seconds \cite{Lambrecht2017,Schaetz2017}, comparable to the lifetime of neutral atoms for similar trapping conditions \cite{Endres2016} and in agreement with theoretical predictions \cite{Cormick2011}. Several groups have also superimposed optical lattices with Coulomb Crystals trapped in rf traps \cite{Laupretre2017,Linnet2012, Karpa2013}, which, e.g., allowed to study fundamental questions in the context of friction \cite{Bylinskii2015}. These experiments were realized providing rf confinement for the radial degrees of freedom, while axial confinement was implemented using optical and electrostatic fields.
\\
In this Letter, we show trapping of up to six ions in a single-beam optical dipole trap without confinement by rf fields. 
We demonstrate that the ensemble remains a one-dimensional Coulomb crystal in the optical trap and reveal access to the axial motional modes.
\begin{figure*}
	  \includegraphics[width=0.95 \textwidth]{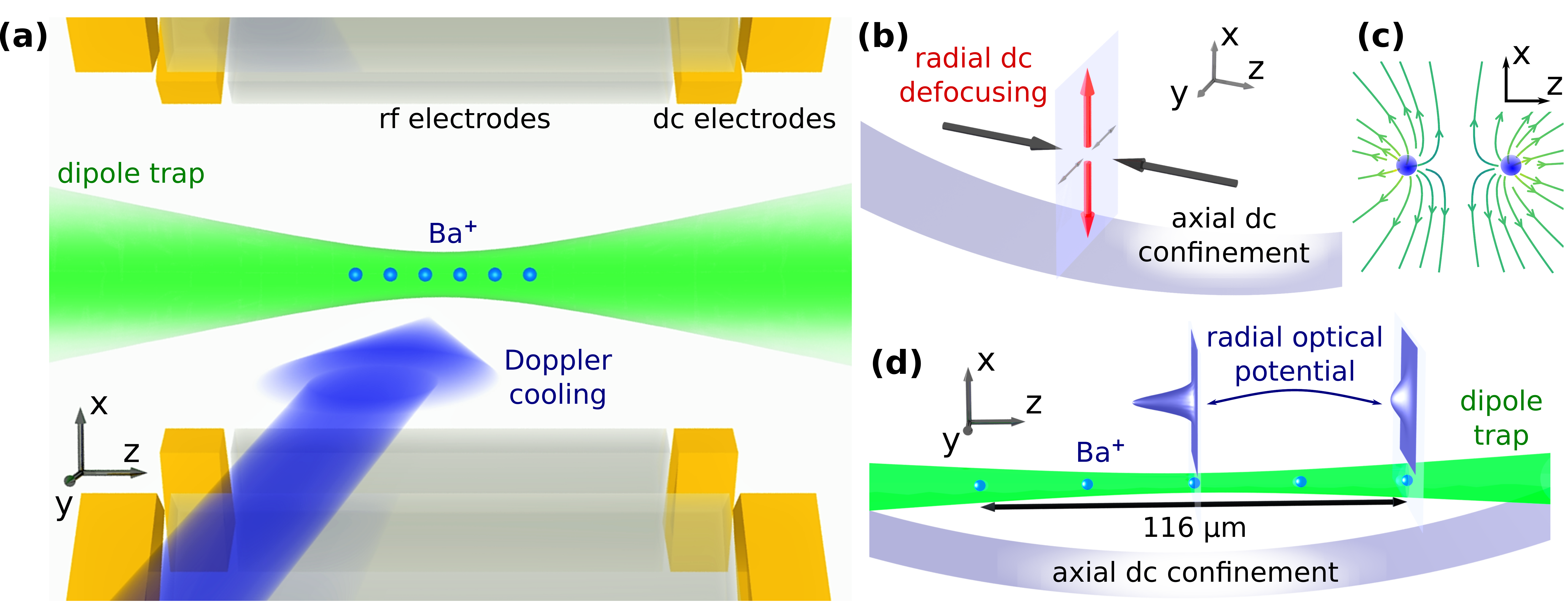}
          \caption{Schematic of the apparatus and contributions to the total trapping potential. 
          (a) We load chains of up to six Doppler-cooled Barium ions, with lengths of up to $135\,\upmu\text{m}$ (5 ions: $116\,\upmu\text{m}$), into a linear rf trap. 
          dc electrodes are used to adjust the external electrostatic potential.
          The ions are then transferred from the rf trap into either a visible (VIS) or near-infrared (NIR) optical dipole trap by ramping up the optical potential and simultaneously turning off the rf fields.
          Axial dc confinement in the external electrostatic potential [grey surfaces in (b) and (d)] as well as the ions' mutual Coulomb interaction (c) lead to additional defocusing forces in the radial directions.
          In our experiment, defocusing is chosen to predominantly lie in the $x$ direction [red arrow in (b)] while defocusing in the $y$ direction is negligible (thin gray arrow).
          With Rayleigh lengths $z_R^\text{VIS} = 40\,\upmu\text{m}$ and $z_R^\text{NIR} = 74\,\upmu\text{m}$ for the respective dipole laser beams, the radial optical potential depicted by the blue surfaces in (d) also depends on the ion position along the $z$-axis.
 \label{fig:TrapSchematic}}
\end{figure*}
\section{\Rmnum{2}. Apparatus for optical trapping of ion Coulomb crystals}
\begin{table}[b]
\begin{ruledtabular}
\begin{tabular}{lrr}
\textbf{Laser} & \textbf{VIS} & \textbf{NIR} \\
$\lambda$ & $532\,\text{nm}$ & $1064\,\text{nm}$\\
$w$ & $(2.6 \pm 0.2)\,\upmu\text{m}$ & $(5.0 \pm 0.2)\,\upmu\text{m}$\\
$z_R$ & $(40 \pm 2)\,\upmu\text{m} $ & $(74 \pm 1)\,\upmu\text{m} $\\
$P_{\text{opt}}$ & $\leq 9.5\,\text{W}$ & $\leq 20\,\text{W}$\\
$U_{\text{opt}}/k_B$ & $\leq (110\pm18)\,\text{mK}$ & $\leq (16 \pm 1)\,\text{mK}$\\
$\omega_{\text{rad,opt}}/(2\pi)$ & $\leq (315\pm25)\,\text{kHz}$ & $\leq (62\pm2)\,\text{kHz}$\\
$\Delta_{\text{Stark}}/(2\pi)$ & $\leq (2.4\pm0.4) \,\text{GHz}$ & $\leq (330\pm30) \,\text{MHz}$\\
\end{tabular}
\end{ruledtabular}
\caption{\label{tab:lasers} Laser and trap parameters for visible (VIS) and near-infrared (NIR) optical dipole traps. 
The symbols denote wavelength ($\lambda$), $1/e^2$ beam waist ($w$), Rayleigh length ($z_R$), laser power ($P$), optical trap depth ($U_{\text{opt}}$), radial trap frequency ($\omega_{\text{rad,opt}}$), and Stark shift in the electronic ground state $S_{1/2}$ ($\Delta_{\text{Stark}}/(2\pi)$).
$\omega_{\text{rad,opt}}$ denotes the radial optical trap frequency (neglecting electrostatic fields) at the mimimum beam waist and for a single ion.}
\end{table}%
Our experimental setup combines focused dipole laser beams with a segmented linear rf trap, see Fig.~\ref{fig:TrapSchematic}(a).
We follow a three step protocol for optical trapping of multiple ions in the absence of rf fields, during which we maintain control over the axial confinement by electrostatic fields.
In step one, we load $1\leq N_{\text{ini}} \leq 6$ ions by photoionization of Barium atoms \cite{Leschhorn2012} and prepare them in the rf trap (drive frequency $\Omega_{\mathrm{rf}}/(2\pi)=1.4\,\text{MHz}$).
The Doppler cooling laser, which is also responsible for detection, cools the gaseous ensemble below the transition to the crystalline phase (cooling rate $\approx 10^3 s^{-1}$, Doppler limit $T_D \approx 0.3\,\text{mK}$).
For temperatures $T$ with $T_D<T\ll T_c$, the chosen trapping frequencies $\omega_{\text{rad}}^{\mathrm{rf}} = 2 \, \pi \times 140\,\text{kHz}$ and  $\omega_{\text{ax}}^{\mathrm{dc}} = 2 \, \pi \times
25\,\text{kHz}$, and $N_{\text{ini}} < 10$, the Coulomb crystals extend as
one-dimensional chains along the $z$-axis of the rf trap with an inter-ion distance
of $\sim 35\,\upmu\text{m}$.
This step includes the compensation of stray electric fields to the
level of $|\vec{E}_{\text{stray}}| \lesssim 10^{-2}\,\text{V/m}$ for
a single ion \cite{Huber2014,Lambrecht2017} and detection by
fluorescence imaging at $493\,\text{nm}$ on the $S_{1/2}$-$P_{1/2}$ transition (natural linewidth $\Gamma=2 \pi \times 15.5$\,MHz) with a CCD exposure time of $ 300 \, \mathrm{ms} $.
While fluorescence imaging can directly resolve $^{138}\text{Ba}^{+}$, admixed $^{136,137}\text{Ba}^{+}$ ions remain dark due to their isotopic shifts being large compared to $\Gamma$.
However, as a consequence of sympathetic cooling and Coulomb repulsion, dark ions can be embedded at random sites of the lattice formed together with the bright $^{138}\text{Ba}^{+}$ ions.
From the dark gaps on the fluorescence images [Fig.~\ref{fig:trappedcrystals}(a)], we deduce the number and configuration of bright ($N_b$) and dark ($N_d$) ions with a fidelity close to one.\\
In step two, the ions are transferred into the optical trap by turning on either the visible (VIS) or near-infrared (NIR) dipole trap while ramping the rf field to zero \cite{Schneider2010,Huber2014,Lambrecht2017}.
The laser and optical dipole trap parameters are shown in Table~\ref{tab:lasers}.
The axial confinement is controlled by dc voltages applied to the endcap electrodes (yellow electrodes in Fig. \ref{fig:TrapSchematic}(a)) which remain unchanged for the remaining protocol.
Both dipole traps are generated by focusing circular Gaussian beams (see Fig.\,\ref{fig:TrapSchematic}) with their wave vector aligned with the linear ion chain centered at the minimal beam waist.
After a duration $\Delta t_{\text{opt}}$, we turn on the rf trap while turning off the dipole trap. \\
In step three, we detect the number $N_{\text{opt}}$ and analyze the configuration of the remaining ions.
The optical trapping probability $p_{\text{opt}}$ is defined as the number of successful trapping attempts divided by total trapping attempts. We call an attempt successful if the number of ions before and after optical trapping is equal, $N_{\text{opt}} = N_{\text{ini}}$, and unsuccessful if one or more ions have been lost, $N_{\text{opt}}<N_{\text{ini}}$. 
In such cases, we find that typically only one or two ions are missing after $\Delta t_{\text{opt}}$.
The statistical uncertainty of $p_{\text{opt}}$ is determined by the Wilson score interval \cite{Wilson1927}.
\section{\Rmnum{3}. Keeping Coulomb order}
\begin{figure}
 \includegraphics[width = 0.48 \textwidth]{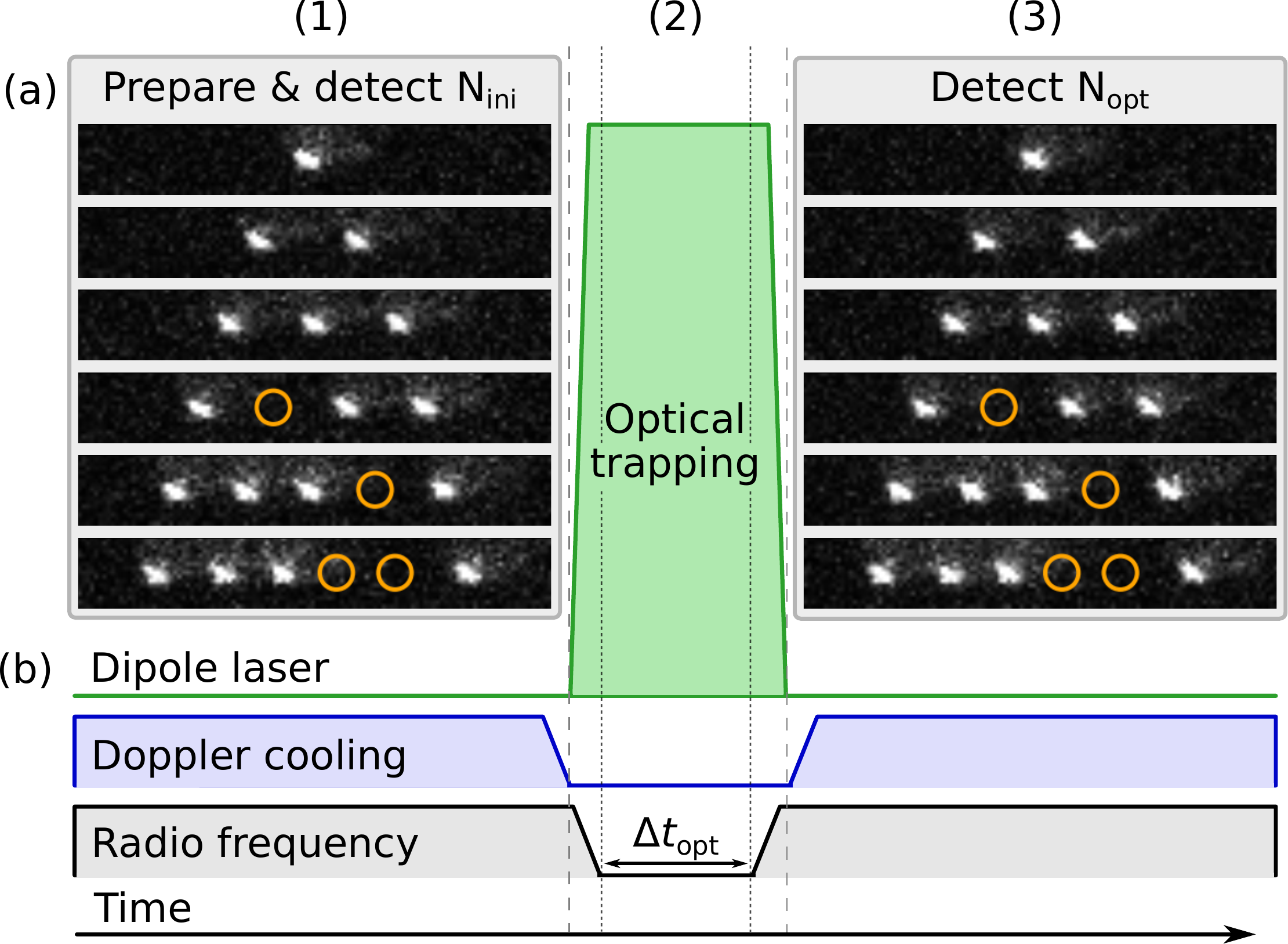}
 \caption{
  Demonstrating the persistence of Coulomb order for an increasing number of optically trapped ions. (a) Fluorescence images of Coulomb crystals with $N_{\text{ini}} = 1...6$ $\text{Ba}^+$ ions are recorded before and after optical trapping of $N_{\text{opt}}$.
  For $ N_{\mathrm{ini}} \ge 4$, the gaps marked by orange circles reveal the presence of dark ions which appear at initial random lattice sites after Doppler and sympathetic cooling.
  (b) The experimental protocol (not to scale) consists of three steps: (1) we detect the initial configuration and ion number $N_{\mathrm{ini}}$ while Doppler cooling the ions;
  (2) the ions are transferred into the dipole trap by turning off the rf field and cooling lasers for the optical trapping duration $ \Delta t_{\mathrm{opt}} $, keeping the electrostatic potential;
  (3) we again detect the number and final configuration of all remaining ions in the rf trap.
  An intermittent gaseous phase followed by recrystallization or enhanced diffusion should be observable with high fidelity via changes of the positions of the dark ions within the crystal.
  \label{fig:trappedcrystals}}
 \end{figure}
For our trapping parameters, theory predicts the existence of one-dimensional ion crystals for temperatures $T < T_c \approx 50\,\text{mK}$, which is on the same order as the available optical trap depth for a single ion, $U_{\text{opt}}^{\text{VIS}} \leq k_B\times 110\,\text{mK}$.
However, this criterion is only valid under certain assumptions, such as homogeneous radial confinement and similar temperatures and heating rates in all spatial degrees of freedom.
These approximations are not fully justified in our experiment.
Therefore, more detailed studies are required to confirm the survival of the crystal during $\Delta t_{\text{opt}}$.\\
We first investigate the feasibility of confining more than one ion in the VIS dipole trap, which provides the deeper potential $U_{\text{opt}}^{\text{VIS}}$. 
However, the VIS trap only creates an attractive potential for $\text{Ba}^+$ in its electronic ground state $S_{1/2}$. In the metastable $D$ states, which can be populated by off-resonant scattering, the potential is repulsive and the ion is lost. 
At $P_{\text{opt}}^{\text{VIS}}=9.5\,\text{W}$, scattering would limit the $1/e$ lifetime of a single ion in the center of the VIS trap to about $1.3 \,\text{ms}$. In order to minimize this effect, we set $\Delta t_{\text{opt}} \approx 500\,\upmu\text{s}$ and turn on additional repumping lasers \cite{Lambrecht2017}.
As shown in Fig.~\ref{fig:trappedcrystals}(a), we demonstrate reliable optical trapping for $N_{\text{ini}} \leq 5$ ions, here permitting $p_{\text{opt}} \approx 1$.
For $N_{\text{ini}} = 6$, we find $p_{\text{opt}} \leq 0.2$.
Note that for $P_{\text{opt}}^{\text{VIS}}=0\;  \text{W}$, no trapping is observed.
Currently, the setup does not allow for direct imaging during $\Delta t_{\text{opt}}$, since the Stark shift inside the VIS laser $\Delta_{\text{Stark}}^{\text{VIS}}$ exceeds $\Gamma$ by orders of magnitude.
The images of the crystals are therefore taken during steps 1 and 3 of the protocol [see Fig.~\ref{fig:trappedcrystals}(a)].
Note that there is no direct evidence that the crystal survived the transfers between the traps. The ensemble might melt and turn into a gas-phase OCP during optical trapping. Then, at the beginning of step 3, it could re-crystalize under the effect of the detection laser and the associated Doppler cooling.\\
To gain deeper insight into the dynamics during transfers and
$\Delta t_{\text{opt}}$, we study the ion ensemble indirectly by embedding $^{136,137}\text{Ba}^+$
as markers to witness changes of the crystalline configuration, see Fig.~\ref{fig:trappedcrystals}(a).
For $N_{\text{ini}} = N_{\text{opt}} \leq 5$ and $N_d \leq 2$, we observe that close to $100\,\%$ of the image pairs show identical configurations of bright and dark ions.
After random reorganization, a given configuration only occurs with a probability of $p_{\text{rand}} = N_{d}!N_{b}!/N_{\text{ini}}!$.
We typically observe that the configurations of 4-ion crystals ($N_{\text{ini}}=4$, $N_d=1$) remain unchanged over the course of 15 consecutive experiments, yielding $(p_{\text{rand}})^{15} = 9 \times 10^{-10}$.
We attribute events with changed configuration ($<1\,\%$ of all image pairs) to collisions of the ions with residual background gas particles during step 1 or 3, leading to melting and recrystallization of the entire ensemble within the deep rf trap.
When a background gas collision occurs during step 2, we expect loss of the ion(s) from the optical trap ($N_{\text{opt}} < N_{\text{ini}}$).
The persistence of Coulomb order is evidence that the thermal excitation of the ensemble remains below $T_c$.
Additionally, the method demonstrates that even isotopes which are not Doppler-cooled (e.g. $^{136,137}\text{Ba}^+$), can be optically trapped when embedded into the ensemble, despite the intrinsically reduced cooling rate.
\section{\Rmnum{4}. Temperature of multiple ions\\ in an optical trap}
\begin{figure}
 \includegraphics[width = 0.48 \textwidth]{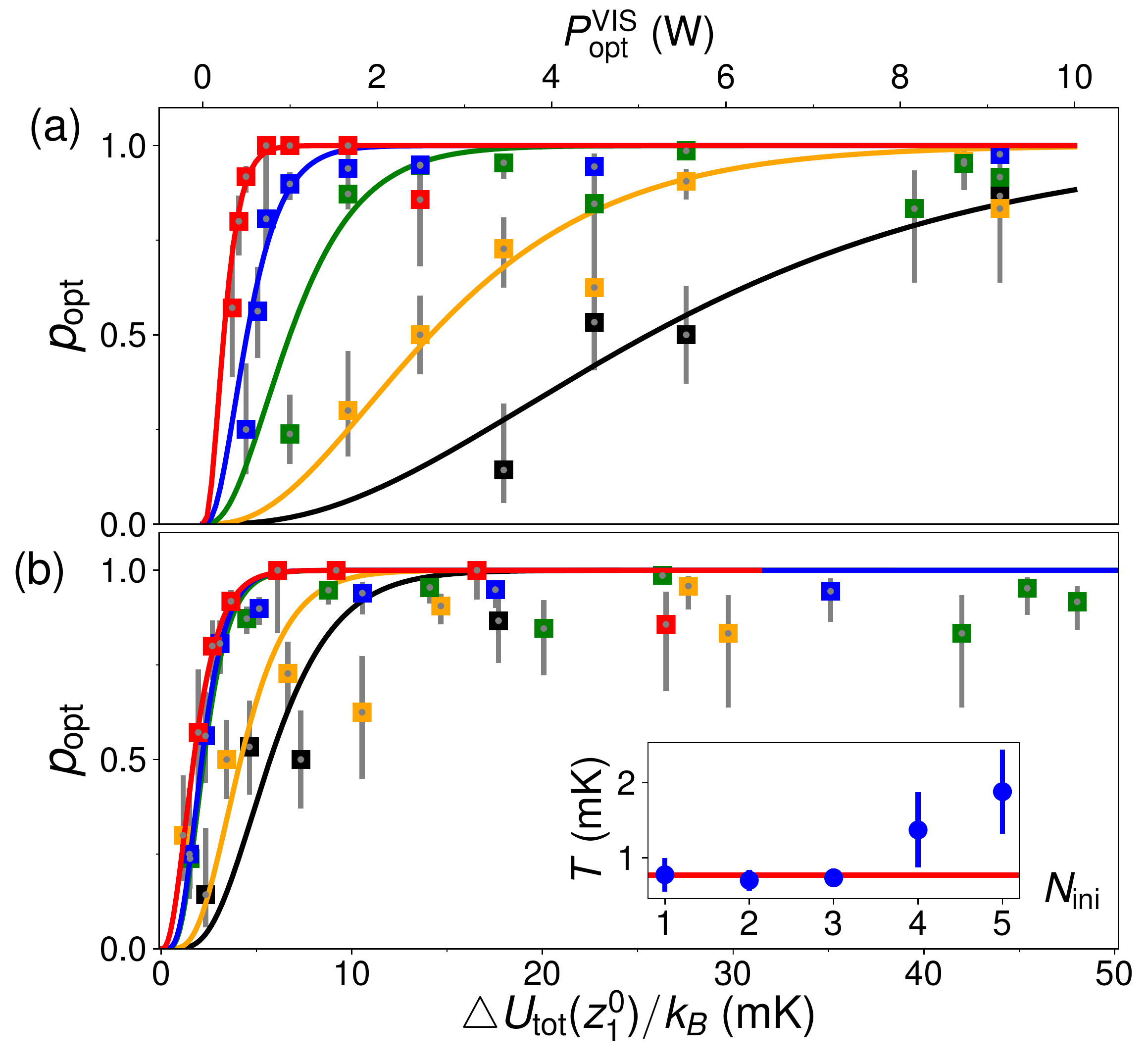}
  \caption{
  Measuring the temperature of multiple ions in the single-beam VIS dipole trap.
  (a) We measure the trapping probability $p_{\text{opt}}$ for $N_{\text{ini}} = 1...5$ ions as a function of $P^{\text{VIS}}_{\text{opt}}$ (red, blue, green, yellow and black squares).
  The solid lines indicate fits with the radial-cutoff model \cite{Schneider2012}, which relates the trapping probability and finite trap depth for atoms or a single ion to their temperature.
  We observe that trapping $N_{\text{ini}}>1$ ions requires increased laser power.
  Still applying the single-particle model yields a ten times larger temperature or decreased trap depth.
  In (b), we reanalyse the data shown in (a), taking into account the axial extent of the ensemble, electrostatic fields and mutual Coulomb interaction to calculate the local radial trap depth at the axial positions of the outermost ions $\Delta U_{\text{tot}}(z^0_{1,N_{\text{ini}}})$.
  Solid lines depict fits assuming the altered radial-cutoff model for $\Delta U_{\text{tot}}(z_i^0)$ (see text).
  We obtain temperatures of $T_{N_{\text{ini}}} = (0.7\pm 0.1) \,\text{mK}$ for $N_{\text{ini}}\leq 3$, as well as $T_4= (1.3 \pm 0.5) \,\text{mK}$ and $T_5= (1.8 \pm 0.5) \,\text{mK}$ (inset).
  \label{fig:temps}
  }
\end{figure}
To access the mean kinetic energy within the ensemble during $\Delta t_{\text{opt}}$ and to study the dominant loss mechanisms, we further investigate the dependence of $p_{\text{opt}}$ on our experimental parameters and measure $p_{\text{opt}}$ for different $P_{\text{opt}}^{\text{VIS}}$.\\
The dependence of $p_{\text{opt}}$ on the optical trap depth $U^{\text{VIS}}_{\text{opt}}$ has previously been exploited to determine the temperature of a single ion \cite{Schneider2010, Huber2014, Lambrecht2017}.
$U^{\text{VIS}}_{\text{opt}}$ defines the cut-off energy for a 3D Boltzmann distribution of indeterminate temperature $T$. 
By integrating the distribution up to $U^{\text{VIS}}_{\text{opt}}$, one obtains the approximate analytic expression $p_{\text{opt}}(\xi) = 1-e^{-2\xi} - 2 \xi e^{-\xi}$, where $\xi = U^{\text{VIS}}_{\text{opt}}/k_B T$, which can be used to derive $T$.\\
Since we have to consider additional effects contributing to the total trapping potential shown in Fig.~\ref{fig:TrapSchematic}, it is evident that this approach is not suitable for $N_{\text{ini}} > 1$, see Fig.~\ref{fig:temps}(a). Taking into account the Gaussian laser beam geometry, mutual Coulomb interaction and the influence of electrostatic fields [see Fig.~\ref{fig:TrapSchematic}(b-d)] makes the radial confinement strongly dependent on $N_{\text{ini}}$ and the axial equilibrium position $z^0_i$ of ion $i$.
These effects are of fundamental importance for optical trapping of charged particles and have to be considered when deriving the local trap depth $\Delta U_{\text{tot}}$ for ion $i$.
First, the ensemble of ions will extend at least up to the length of the Coulomb crystal in the rf trap for the chosen trapping parameters and $T < T_c$.
Because the axial extensions are comparable to the Rayleigh length, e.g. $ |z_1^0 - z_6^0| \approx 135\,\upmu\text{m} > 2 \times  z_{\text{R}}^{\text{VIS}}$ for $N_{\text{ini}} = 6$ [see fig. \ref{fig:TrapSchematic}(d) and table \ref{tab:lasers}], the trap depth depends on the axial position(s) of the ion(s). Even when only considering the optical potential, the increasing laser beam size $w^{\text{VIS}}(z^0_i)$ results in a reduced trap depth $U^{\text{VIS}}_{\text{opt}}(z^0_i)$, see Fig.~\ref{fig:TrapSchematic}(d).
\\
Additionally, the electrostatic potential set by dc electrodes and the mutual Coulomb interaction lead to a potential energy $U_{\text{el}}(\vec{r}_i=(x_i,y_i,z_i)) = U_{\text{dc}}(\vec{r}_i) + U_{\text{coul}}(\vec{r}_i)$. 
This results in effective radial defocusing, which can be seen by expanding $U_{\text{el}}(\vec{r}_i)$ to second order, yielding potential curvatures $m \tilde{\omega}^2_{x,\text{el}}(z^0_i)$ and $m \tilde{\omega}^2_{y,\text{el}}(z^0_i)$ for ion $i$ with mass $m$ near $\vec{r}_i^{\;0}=(0,0,z_i^0)$.
The contribution by the electrostatic potential energy $U_{\text{dc}}(\vec{r}_i)$ expanded to second order defines characteristic potential curvatures $m \tilde{\omega}_{(x,y,z),\text{dc}}^2$, related via the Laplace equation, $\tilde{\omega}_{x,\text{dc}}^2 + \tilde{\omega}_{y,\text{dc}}^2 + \tilde{\omega}_{z,\text{dc}}^2 = 0$. 
Positive terms can be interpreted as trapping frequencies and imply confinement, e.g. in the axial direction $\tilde{\omega}_{z,\text{dc}}^2>0$, whereas negative terms correspond to defocusing, inevitable in at least one radial direction.
In our setup, we find that the defocusing almost exclusively occurs along the $x$ direction (see Fig.~\ref{fig:TrapSchematic}) such that $\tilde{\omega}_{x,\text{dc}}^2 \approx - \tilde{\omega}_{z,\text{dc}}^2$. We therefore neglect the residual defocusing along the $y$ direction, making the $x$ direction the preferred escape path for the ions.
We then approximate the Coulomb interaction $U_{\text{coul}}(\vec{r}_i)|_{z_i = z_i^0}$ of ion $i$ with all other ions, assuming that the ions remain at their equilibrium positions.
This results in an additional defocusing in the $x$ and $y$ directions $\tilde{\omega}_{x,\text{coul}}^2(z_i^0)=\tilde{\omega}_{y,\text{coul}}^2(z_i^0)<0$.
Finally, we approximate the curvature of $U_{\text{el}}(\vec{r}_i)$ via $\tilde{\omega}_{x,\text{el}}^2(z^0_i)= \tilde{\omega}_{x,\text{dc}}^2+\tilde{\omega}_{x,\text{coul}}^2(z_i^0)$ for ion $i$ in the weakest confined direction $x$.
The total radial potential energy for ion $i$ at the axial position $z^0_i$ is then written as $U_{\text{tot}}(x_i,y_i,z_i^0) = U^{\text{VIS}}_{\text{opt}}(x_i,y_i,z_i^0) + U_{\text{el}}(x_i,y_i,z_i^0)$.
In the following, the difference between the local maximum and minimum of $U_{\text{tot}}(x_i,0,z^0_i)$ along the $x$ direction will be referred to as the local radial trap depth $\Delta U_{\text{tot}}(z^0_i)$ (see Supplemental Material for details).
\\
The measured optical trapping probability for $N_{\text{ini}}$ ions and laser power $P^{\text{VIS}}_{\text{opt}}$ is modeled as the product of the individual trapping probabilities of the ions, $p_{\text{opt}}(N_{\text{ini}})=\prod_{i\leq N_{\text{ini}}}{p_{\text{opt,ind}}(\xi_i)}$, where $\xi_i = \Delta U_{\text{tot}}(z^0_i)/k_B T$.
This allows fitting $p_{\text{opt}}(N_{\text{ini}})$ for each $N_{\text{ini}} \leq 5$.
In Fig.~\ref{fig:temps}(b), we show $p_{\text{opt}}(N_{\text{ini}})$ in dependence on the smallest trap depth, $\Delta U_{\text{tot}}(z^0_{1})=\Delta U_{\text{tot}}(z^0_{N_{\text{ini}}})$ at the edges of the ensemble.
We derive temperatures near $T_{N_{\text{ini}}} = (0.7 \pm 0.1) \,\text{mK}$ for $N_{\text{ini}}\leq 3$, $T_4= (1.3 \pm 0.5) \,\text{mK}$ and $T_5= (1.8 \pm 0.5) \,\text{mK}$.
The apparent increase in temperature for larger $N_{\text{ini}}$ may stem from a residual angle between the rf and optical axis, which becomes increasingly relevant for larger ion numbers.
In addition, the trap depth, and therefore $p_{\text{opt}}(N_{\text{ini}})$, is affected by stray fields (which are currently compensated at the position of the center ion only) and deviations from the assumed beam profile for increasing distance $|z_i|$.
Nonetheless, including the spatial dependence of the total radial confinement improves the description of the system and yields $T < 2\,\text{mK} \ll T_c $ during $\Delta t_{\text{opt}}$.
For these temperatures and our experimental parameters, MD simulations show that the amplitudes of the ions' axial motion are small ($<10\,\%$) compared to the distance of neighboring ions.
Thus, applying the Lindemann criterion for a small number of lattice sites, we conclude that ensembles of up to 5 ions form crystals during $\Delta t_{\mathrm{opt}}$ \cite{Lindemann1910, Drewsen2015}.\\
In our system, the number of ions forming a crystal in the VIS trap is limited by our beam geometry and the chosen trapping parameters.
Further improvement could be achieved by adapting the laser beam geometry, replacing axial electrostatic confinement with optical confinement or by using different ion species featuring either a smaller decay rate into repulsive $D$ states or no such states at all.
In the next section, we will use the further detuned NIR optical trap to reduce off-resonant scattering, at the expense of the confinement, see table~\ref{tab:lasers}.
\section{\Rmnum{5}. Detecting motional modes of optically trapped ions}
\begin{figure}[b]
 \includegraphics[width = 0.48 \textwidth]{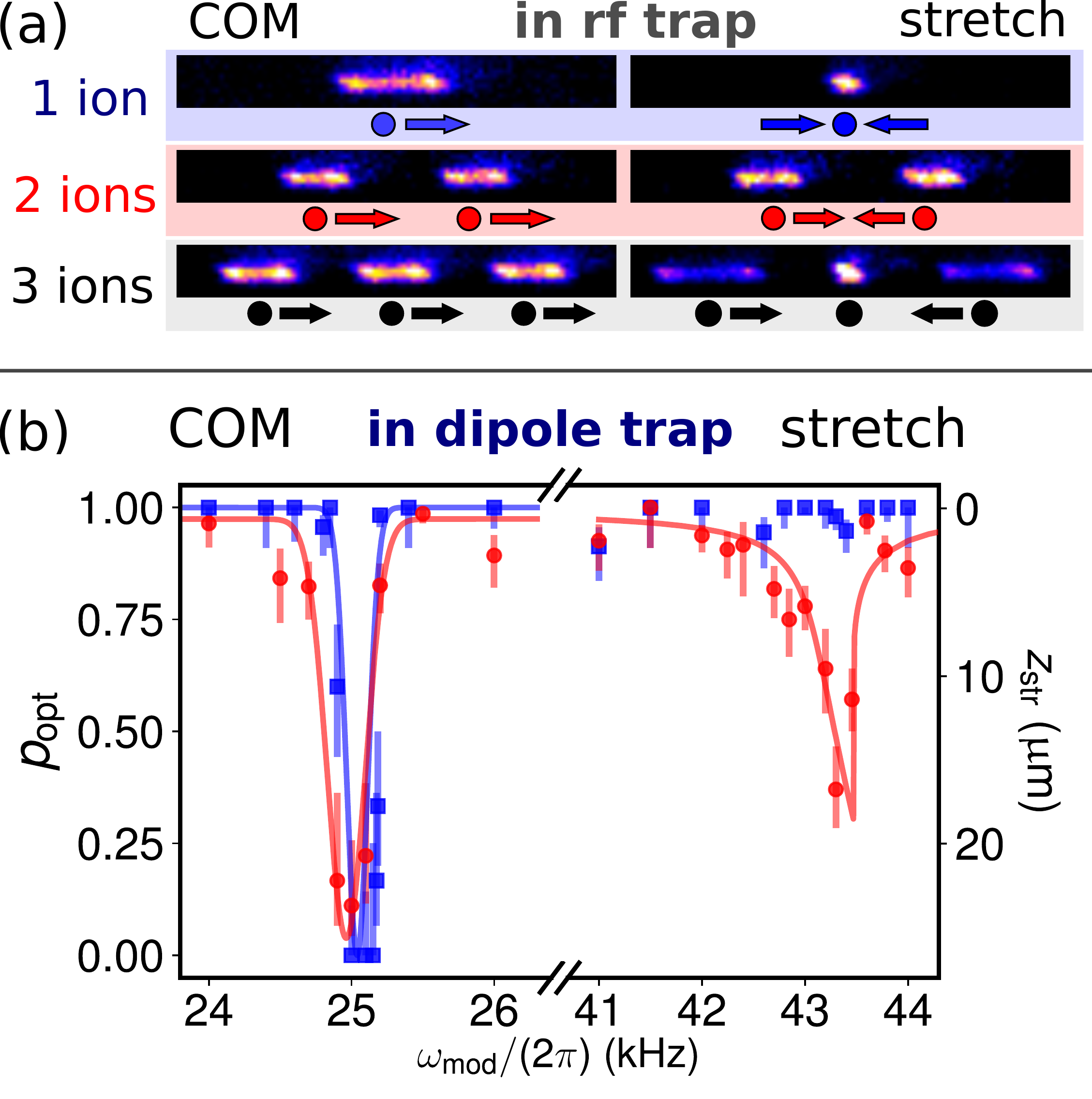}
 \caption{
 Spectrometry of normal modes, demonstrating access to the axial phonons of the crystal during optical trapping.
  In (a), we show typical fluorescence images of ions in the rf trap with dc axial confinement, resonantly modulated with oscillating electric fields (two-ion distance $43\,\upmu\text{m}$, other images to scale).
  For $N_{\text{ini}}=1$, we observe a single resonance for the axial motion at $\omega_{\text{ax}}^{\text{COM}}$.
  For $N_{\text{ini}}=2,3$, an additional resonance at $\omega_{\text{ax}}^{\text{str}}= \sqrt{3}\, \omega_{\text{ax}}^{\text{COM}}$ corresponding to out-of-phase motion appears.
  (b) Optical trapping probability for $N_{\text{ini}}=1$ (blue squares) and $N_{\text{ini}}=2$ (red circles) ion(s) in the NIR trap, as a function of the frequency $\omega_{\text{mod}}/2\pi$ of the oscillating electric field.
  We observe a drop in $p_{\text{opt}}$ at $\omega_{\text{ax}}^{\text{COM}}$ for both $N_{\text{ini}}=1$ and $N_{\text{ini}}=2$, in agreement with the expected axial confinement.
  For the COM mode, the solid lines show fits to the data.
  The resonance at $\omega_{\text{ax}}^{\text{str}} = 2\pi \times (43.3\pm0.15)\, \text{kHz}\approx \sqrt{3} \, \omega_{\text{ax}}^{\text{COM}}$ (binned data points weighted with their statistical significance, for details on uncertainty see text) only emerges in the case $N_{\text{ini}}=2$ and shows access to the motional degrees of freedom of the Coulomb crystal during $\Delta t_{\text{opt}}$.
  In the case of the stretch mode, we numerically simulate the out-of-phase motion driven by an oscillating electric field with amplitude $E=1.8\,\text{mV}/\text{m}$ (no free parameters).
  We depict the amplitude $|z_{\text{str}}| = |z_1-z_2|/2$ of this motion after $\Delta t_{\text{opt}}=10\,\text{ms}$ by the solid red line (axis on the right-hand side).
  The nonlinearity of the Coulomb interaction leads to an asymmetric frequency response (see Supplemental Material).
  We emphasize that the dependence of $p_{\text{opt}}$ on $|z_{\text{str}}|$ is nontrivial and not taken into account here.
  \label{fig:spectrum}}
\end{figure}
To gain further insight into the dynamics of the ensemble during $ \Delta
t_{\mathrm{opt}}$, we investigate the vibrational spectrum
of the ions.
Here, we choose to explore the axial degree of freedom.
Coupling to the charge allows exciting the motion of trapped ions by applying oscillating voltages to specific electrodes. 
Since the position of the ions in the rf trap can be observed directly on the CCD camera, we observe motional excitation as an effectively increased ion image size caused by the integration of the fluorescence of the ion along its trajectory [see Fig.~\ref{fig:spectrum}(a)]. 
The blurring and its dependence on the frequency of excitation, $\omega_{\text{mod}}$, allows to calibrate the parameters of the trapping potential. 
The motion of more than one ion forming a Coulomb crystal is typically described in terms of normal modes \cite{Wineland2013}.
In Fig.~\ref{fig:spectrum}(a), the collective motion in the rf trap is presented exemplarily for two axial modes of up to three ions.
The mode of lowest axial frequency, the center-of-mass (COM) mode, describes the in-phase oscillation of all ions at $\omega_{\text{ax}}^{\text{COM}}$. 
The next higher frequency at $\omega_{\text{ax}}^{\text{str}} = \sqrt{3}\, \omega_{\text{ax}}^{\text{COM}}$ corresponds to the
stretch mode, where two ions oscillate opposite in phase. 
For this mode and odd ion numbers, the center ion has to remain at rest.\\
To perform motional spectrometry of ions in the optical trap, we repeat the protocol shown in Fig.~\ref{fig:trappedcrystals}(b) for $N_{\text{ini}}=1,2$ and apply an oscillating voltage of constant amplitude to the endcap electrodes [see Fig.~\ref{fig:TrapSchematic}(a)] during $ \Delta t_{\mathrm{opt}}$ only.
We choose $\Delta t_{\text{opt}} = 10\,\text{ms}$ within the NIR trap to improve the frequency
resolution while further mitigating off-resonant scattering to $\leq 10\,\text{s}^{-1}$.
At specific $\omega_{\text{mod}}$, we observe reduced optical trapping probabilities. 
We identify these frequencies as resonances within the oscillation spectrum of the ensemble during $\Delta
t_{\mathrm{opt}}$. For $N_{\text{ini}}=1$, we observe a single resonance, centered at  $\omega_{\text{ax}}^{\text{COM}}=2\pi \times(25.04\pm0.02)\,\text{kHz}$.
For $N_{\text{ini}}=2$, in addition to the resonance at
$\omega_{\text{ax}}^{\text{COM}}=2\pi \times (24.96\pm0.02)\,\text{kHz}$, we find another pronounced drop of $p_{\text{opt}}$ centered at $\omega_{\text{ax}}^{\text{str}}=2\pi \times (43.3 \pm 0.15)\,\text{kHz} \approx \sqrt{3}\, \omega_{\text{ax}}^{\text{COM}}$.
The experimental uncertainty of $\omega_{\text{ax}}^{\text{str}}$ is estimated via the spacing between adjacent data points which is consistent with the frequency resolution of the excitation.
We interpret the additional resonance as the spectrometric fingerprint of the stretch mode. 
Based on the agreement with the theoretical prediction we confirm the survival of the crystal in absence of any rf while demonstrating the possibility to address and exploit its normal modes.\\
To study the dynamics of the motional excitation and related loss mechanisms, we compare our experimental results with numerical simulations for $N_{\text{ini}}=1,2$ (for details see Supplemental Material). 
The results are in agreement with our experimental findings and reveal that the motional amplitudes along the z-axis lead to a loss of ion(s) in radial direction.
In the case of the COM mode, loss is dominated by the reduction of the optical radial confinement for larger axial displacement [Fig.~\ref{fig:TrapSchematic}(d)], while in the case of the stretch mode, it is caused by the enhanced mutual Coulomb repulsion during the phase of approximation and the forced evasion of the ions in radial direction [Fig.~\ref{fig:TrapSchematic}(c)].
The non-linearity induced by the latter explains the directly
observable anharmonicity of the resonance for the case of the
stretch mode.\\
\section{\Rmnum{6}. Conclusions and outlook}

In summary, we demonstrate trapping of Coulomb crystals in a single-beam optical trap. 
We reveal the importance of Coulomb interaction and the electrostatic field along the axis of the dipole trap and demonstrate access to the collective motion. 
For neutral particles, these effects can usually be neglected.
On the one hand, Coulomb interaction establishes the axial and radial motional modes in ion crystals. 
Control of these modes on the single-phonon level in optical traps would permit to couple electronic degrees of freedom of the ions. 
As in rf traps, phonons could mediate spin-spin interaction between the ions and act as a data-bus \cite{Wineland2013}. In addition, they can feature as quasi-particles which span the bosonic degree of freedom in open quantum systems \cite{Clos2016} and extend experimental quantum simulations, e.g. by tunneling between the lattice sites defined by the ions \cite{Schneider2012b,Bermudez2012}.
On the other hand, ion interaction and electrostatic forces currently limit the size of optically trappable Coulomb crystals.
These contributions modify the trapping potential itself and reduce its depth. 
We aim to increase the number of ions and the dimensionality of the crystals using the range of readily available dipole trap geometries, e.g. Bessel beams, optical lattices or additional (crossed) laser beams.\\
We also embed sympathetically cooled ions, here Barium isotopes, without substantially affecting the temperature of the crystal. 
Given suitable electronic transitions and sufficient coupling strengths, it is possible to optically trap ions of different electronic states and exploit the state-dependent potential \cite{Lambrecht2017}. 
Co-trapping other ionic species and molecular ions \cite{Drewsen2015} should also be considered.\\
We argue that systems in which rf micromotion exists due to intrinsic displacement from the center of the trap, as in higher-dimensional Coulomb crystals or during the interaction with (cold) neutral atoms, could substantially benefit from optical trapping of ions.
While still in its infancy, the technique presented here could provide a clean platform to experimentally investigate systems with predicted quantum phase transitions and feature quantum many-body effects, briefly described in the following.\\
The number of ions and the ratio of radial and axial confinement determine whether a crystal exists in a 1D chain or 2D zigzag structure. 
The two symmetric configurations of zigzag and ``zagzig'' are trapped within an effective double-well potential with well-controllable barrier height and are predicted to allow for experimental studies of a wide range of physical effects, starting with $N\geq 3$. 
Adiabatically reducing the radial confinement to cross the structural quantum phase transition from 1D to 2D has been proposed to create a superposition of zigzag and zagzig \cite{Retzker2008}.
The impact of quantum fluctuations at criticality is predicted to dominate the structure adopted by ions cooled close to the motional ground state, that is, even at finite temperatures \cite{Shimshoni2011}.
It has also been proposed to create an entangled state, incorporating both structural phases, linear and zigzag, simultaneously \cite{Simshoni2011b,Baltrusch2011,Baltrusch2012}. 
Preparing one ion of a linear chain in a coherent superposition of two electronic states and exploiting the state-dependence of the optical trap \cite{Lambrecht2017} could directly implement this proposal.\\
Additionally, embedding a single ion in a BEC has been proposed as a controlled quantum many-body system driven by the nucleation of tens or hundreds of atoms polarized in the ion's electric field \cite{Cote2002,Schurer2017}. 
In hybrid traps, which combine an rf trap for the ion with an optical trap for the atoms, micromotion limits sympathetic cooling of the ion to a regime above the ultracold temperatures required for the formation the clusters \cite{Cetina2012, Meir2016, Tomza2017}. 
Optical trapping of ions may be a generic solution to overcome this limitation \cite{Tomza2017} even for $N>1$ and higher-dimensional structures. 
Additional ions located outside the cloud could act as a remote sensor of the ion-atom interaction. 
It has also been proposed to immerse a linear ion chain into a degenerate Fermi gas to, e.g., emulate solid-state physics, with atoms acting as electrons and ions as nuclei, or to study a Peierls-like phase transition \cite{Bissbort2013,Gerritsma2012}.
Recently, envisioning a setup similar to our current experimental realisation, it has been proposed to couple an array of particles ($N\geq 2$) by coherent scattering inside a light field without an optical cavity \cite{Ostermann2016}.

\section{Acknowledgements}

We thank J.~Denter for technical support. This project has received funding from the European Research Council (ERC) under the European Union's Horizon 2020 research and innovation program (grant agreement n$^\circ$ 648330). J.S., A.L., P.W. and M.D. acknowledge support from the DFG within the GRK 2079/1 program. P.W.  gratefully acknowledges financial support from the Studienstiftung des deutschen Volkes. L.K. is grateful for financial support from Marie Curie Actions.
\section{Supplemental material}
\textbf{Calculation of the local radial trap depth $\Delta U_{\textnormal{tot}}(z^0_i)$:}
\begin{figure}
 \includegraphics[width = 0.48 \textwidth]{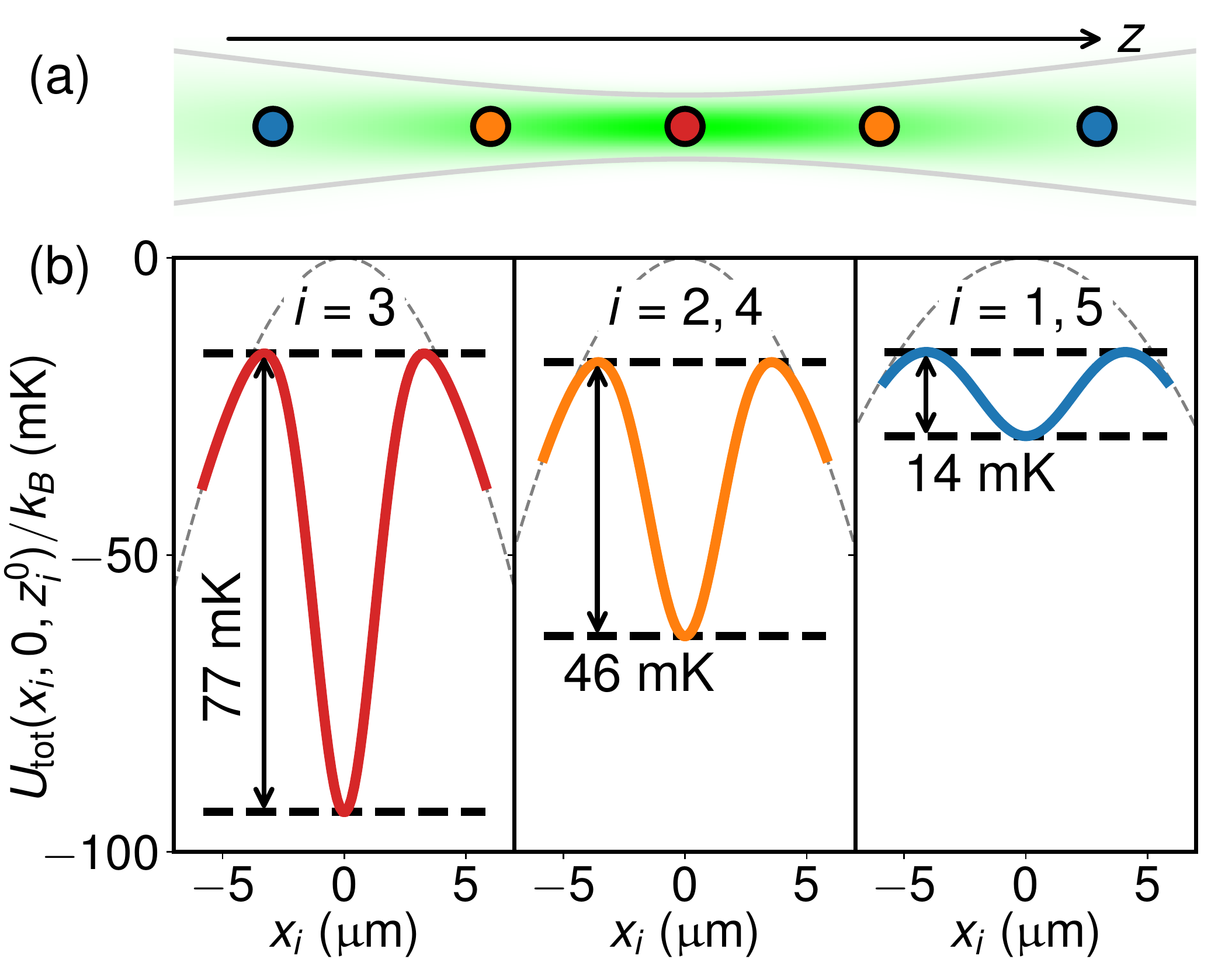}
 \caption{ (a) Sketch of a five ion crystal and (b) trap depths for the optical and electrostatic potential at the equilibrium positions $z_i^0$ of the individual ions.
 The ions with index $i=1...5$ are represented by the blue, orange and red circles and experience different optical potentials depending on their axial position along $z$.
 The optical potential is calculated assuming a Gaussian laser beam with power $P^{\text{VIS}}_{\text{opt}}=8\,\text{W}$ and minimal beam waist radius $w^{\text{VIS}}_x=2.6 \, \upmu\text{m}$.
 The electrostatic potential $U_{\text{el}}(\vec{r}_i)$ (dashed gray line) is approximated as a quadrupole potential and consists of an external electrostatic potential $U_{\text{dc}}(\vec{r}_i)$ with maximum defocusing in the $x$ direction and mutual Coulomb interaction $U_{\text{coul}}(\vec{r}_i)$ which leads to the strongest defocusing for the center ion.
 The total potential is then evaluated at the position of the ions with $i=3$ (red), $i=2,4$ (orange) and $i=1,5$ (blue).
 Note the offset of the local maxima from zero which is due the defocusing given by $U_{\text{el}}(\vec{r}_i)$.\label{fig:trapdepths}}
\end{figure}
We calculate the local radial trap depth at the position of the individual ions. 
Ions are intrinsically sensitive to electric fields due to their charge. 
When applying axial electrostatic confinement $U_{\text{dc}}(\vec{r}_i)$ and taking into account mutual Coulomb interaction $U_{\text{coul}}(\vec{r}_i)$, Laplace's equation predicts defocusing in at least one radial direction.
To derive the ions' trap depth in the absence of rf fields, we have to consider $U_{\text{el}}(\vec{r}_i) = U_{\text{dc}}(\vec{r}_i) + U_{\text{coul}}(\vec{r}_i)$.
Only taking into account second-order terms (stray field compensation makes the first-order terms negligible) leads to a quadrupole potential $U_{\text{el}}(\vec{r}_i) = m/2\, \left(\tilde{\omega}_{x,\text{el}}^2 x_i^2 + \tilde{\omega}_{y,\text{el}}^2 y_i^2 +\tilde{\omega}_{z,\text{el}}^2 z_i^2 \right)$ with $\tilde{\omega}_{x,\text{el}}^2 + \tilde{\omega}_{y,\text{el}}^2 + \tilde{\omega}_{z,\text{el}}^2 = 0$ where $m$ denotes the mass of the ion.
Coulomb interaction and radial optical potential depend on the axial equilibrium position $z_i^0$.
For the defocusing of $U_{\text{el}}(\vec{r}_i)$ in $x$-direction, represented by $\tilde{\omega}^2_{x,\text{el}}(z_i^0)<0$, a Gaussian laser beam with radius $w(z_i^0)$ along $x$ and optical trap depth $U_{\text{opt}}(z_i^0)<0$, the total potential along $x$ is given by
\begin{equation}
 U_{\text{tot}}(x,0,z_i^0) = \frac{1}{2} m \tilde{\omega}_{x,\text{el}}^2(z_i^0)\, x^2 + U_{\text{opt}}(z_i^0)\, \exp{\frac{-2x^2}{w^2(z_i^0)}} \, .
\end{equation}
This function is plotted in Fig.~\ref{fig:trapdepths} for typical experimental values.
The locations of the local minima and maxima are determined by
\begin{align}
 &x_{i,\text{min}}=0 \quad \text{and}\\
 &x_{i,\text{max}}^2(z_i^0)=\frac{w^2(z_i^0)}{2} \, \log{\left(\frac{ 4U_{\text{opt}}(z_i^0)}{m \tilde{\omega}_{x,\text{el}}^2(z_i^0)\, w^2(z_i^0)}\right)} \, .
\end{align}
 Calculating the trap depth as $\Delta U_{\text{tot}}(z_i^0) = U_{\text{tot}}(x_{i,\text{max}},0,z_i^0)-U_{\text{tot}}(x_{i,\text{min}},0,z_i^0)$ yields the following expression:
\begin{equation}
\begin{split}
\Delta U_{\text{tot}}(z_i^0) = - U_{\text{opt}}(z_i^0) &+ \\
+ \frac{m \tilde{\omega}_{x,\text{el}}^2(z_i^0)\, w^2(z_i^0)}{4} &\left[1 + \log\left(\frac{4 U_{\text{opt}}(z_i^0)}{m \tilde{\omega}_{x,\text{el}}^2(z_i^0)\, w^2(z_i^0)} \right)\right].
\end{split}
\end{equation}
The second term in this expression, which corresponds to the offset of $U_{\text{tot}}(\pm x_{i,\text{max}},0,z_i^0)$ from zero visible in Fig.~\ref{fig:trapdepths}(b), is only well defined when $m \tilde{\omega}_{x,\text{el}}^2(z_i^0)\, w^2(z_i^0) < 4 U_{\text{opt}}(z_i^0)$.
This condition is violated when the electrostatic defocusing is stronger than the optical confinement, prohibiting the existence of a local minimum at $x=0$ such that $\Delta U_{\text{tot}}(0,0,z_i^0) =0$.

\textbf{Numerical simulations of the normal mode detection:}
In the following, we discuss in more detail the results of the numerical model shown in Fig. \ref{fig:spectrum}(b), depicting the motional amplitude of the axial out-of-phase (stretch) mode for large excitation amplitudes.
\begin{figure}[h!]
 \includegraphics[width = 0.48 \textwidth]{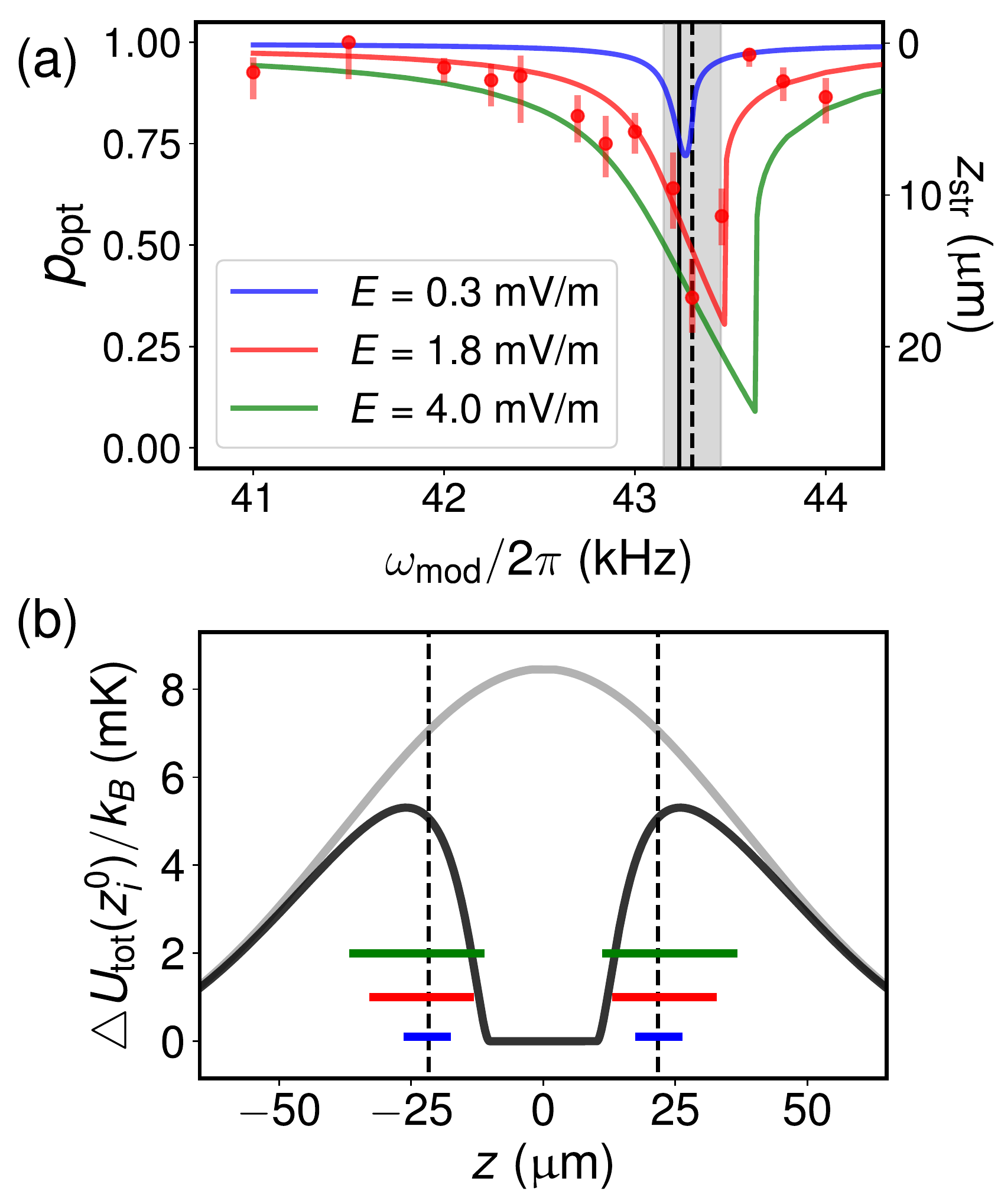}\\
 \caption{Behavior of the ions' out-of-phase oscillation in the optical trap. (a) Numerical simulation of the frequency response of the out-of-phase oscillation (solid lines) and experimental results for spectrometry of two ions, also shown in Fig.~\ref{fig:spectrum}. 
 The simulation is repeated for different amplitudes of the driving field, $E= (0.3,1.8,4.0)\, \text{mV/m}$ for the (blue, red, green) curves.
 The solid vertical line corresponds to $\sqrt{3}\,\omega_{\text{COM}}$, the expected position of $\omega_{\text{str}}$ for the measured resonance of the in-phase motion for two ions $\omega_{\text{COM}}$.
 The dashed vertical line is the position of the minimum of $p_{\text{opt}}$ and the gray shaded area corresponds to its estimated uncertainty.
 The measured frequency and the asymmetric shape of the data are in excellent agreement with the simulation results (no free parameters).
 (b) Radial trap depths for one ion (gray line) and two ions (black line) as a function of axial position $z$ and trajectories of ions for varying driving amplitudes (blue, red and green horizontal lines).
 For a single ion, the trap depth is given by the combination of an electrostatic and an optical potential and is largest at $z=0$ due to divergence of the beam.
 For two ions at $z_{1,2}$ (equilibrium positions $z_{1,2}^0$ depicted by vertical black dashed lines), the additional defocusing from Coulomb interaction lowers the trap depth as the ions approach each other, e.g. when oscillating at large amplitudes and out-of-phase.
 The lowered trap depth radially expels the ions from the optical trap.
 Due to the anharmonicity of the potential, the oscillation is asymmetric.
 \label{fig:excite_str}}
\end{figure}
The axial potential is dominated by electrostatic forces which provide near-harmonic confinement, and Coulomb interaction leads to equilibrium positions $z_{1,2}^0$.
In the experiment, we drive the system with an oscillating electric field $E(t) = E\sin{(\omega_{\text{mod}}t)}$ by applying voltages to the endcap electrodes.\\
Now considering the time-dependent positions $z_{1,2}(t)$ of the two ions, we obtain a system of coupled harmonic oscillators with nonlinear coupling given by Coulomb interaction $\mp e^2 / 4 \pi \epsilon_0 |z_1(t) - z_2(t)|^2$ (negative for $z_1$ and positive for $z_2$ when $z_1<0<z_2$).
Damping can be neglected as laser cooling is inactive during optical trapping.
The driving terms for $z_1$ and $z_2$ have the form $+eE \sin{(\omega_{\text{mod}}t)}$ and $\pm eE \sin{(\omega_{\text{mod}}t)}$, where the identical signs correspond to a homogeneous field used to excite the COM mode and the opposite signs correspond to a field gradient used to excite the stretch mode.
By adding or subtracting the equations of motion and defining $z_{\text{COM}}=(z_1 + z_2)/2$ and $z_{\text{str}}=(z_1 - z_2)/2$, we obtain the decoupled equations:
\begin{equation}
 \ddot{z}_{\text{COM}}(t) + \omega_z^2 z_{\text{COM}}(t) - \frac{eE}{m} \sin{(\omega_{\text{mod}}t)} = 0 \label{eq_diffeq1}
\end{equation}
and
\begin{equation}
 \ddot{z}_{\text{str}}(t) + \omega_z^2 z_{\text{str}}(t) + \frac{e^2}{8 \pi \epsilon_0 m |z_{\text{str}}(t)|^2} - \frac{eE}{m} \sin{(\omega_{\text{mod}}t)} = 0 \label{eq_diffeq2} \, .
\end{equation}
The out-of-phase motion behaves like a harmonic oscillator for small amplitudes, when the change of the distance between the ions, and thereby their Coulomb interaction, is small.
For large driving amplitudes, we solve the differential equation (\ref{eq_diffeq2}) numerically for a varying frequency $\omega_{\text{mod}}$, see Fig. \ref{fig:excite_str} (a).\\
At $t=0$, the ions are assumed to be at rest at their equilibrium positions in a harmonic potential described by the measured frequency of the two-ion COM mode $\omega_{\text{ax}}^{\text{COM}} = 2 \pi \times (24.96\pm 0.02) \, \text{kHz}$.
We then let the system evolve for $\Delta t = 10\,\text{ms}$ (time step $\delta t = 1 \,\upmu\text{s}$).
The oscillation amplitude is defined as half the distance between the maximal and minimal separation of the ions $\text{max}(z_{\text{str}}(t))-\text{min}(z_{\text{str}}(t))$.
The asymmetric shape of the resonance appears when the ions are sufficiently close to experience modified Coulomb interaction.
Additionally, the largest oscillation amplitude is shifted to higher frequencies with increased driving amplitude.
In the experiment, the amplitude of the driving electric field is chosen as $E = (1.8 \pm 0.1)\,\text{mV/m}$.
To investigate the sensitivity of this parameter, we vary $E$ in the simulation, see Fig. \ref{fig:excite_str}(a).\\
The detection of the oscillation is achieved by measuring the trapping probability.
This raises the question how the ions are lost from the trap.
As the axial position changes during the resonant excitation, the radial optical trap depth changes as well.
We distinguish two possible loss mechanisms.
(a) If the ions move away from the focus of the laser beam, the beam waist increases and the radial trap depth approaches zero. 
This may lead to loss when exciting the COM or stretch mode.
(b) When resonantly exciting the stretch mode, the ions perform out-of-phase motion.
When their distance is minimal, Coulomb repulsion leads to stronger radial defocusing, reducing the trap depth.
More intuitively, the effect can be described as a ``collision'' of the ions in the guide created by the optical trap.
When this happens, they avoid each other, moving to the side and leaving the trapping region.
In figure \ref{fig:excite_str}(b), the axial oscillation amplitude of the out-of-phase motion is shown together with the dependence of the radial trap depth on the ion distance $d$.
According to this calculation, the ions leave the trap through the second loss mechanism.

\begin{thebibliography}{45}%
\makeatletter
\providecommand \@ifxundefined [1]{%
 \@ifx{#1\undefined}
}%
\providecommand \@ifnum [1]{%
 \ifnum #1\expandafter \@firstoftwo
 \else \expandafter \@secondoftwo
 \fi
}%
\providecommand \@ifx [1]{%
 \ifx #1\expandafter \@firstoftwo
 \else \expandafter \@secondoftwo
 \fi
}%
\providecommand \natexlab [1]{#1}%
\providecommand \enquote  [1]{``#1''}%
\providecommand \bibnamefont  [1]{#1}%
\providecommand \bibfnamefont [1]{#1}%
\providecommand \citenamefont [1]{#1}%
\providecommand \href@noop [0]{\@secondoftwo}%
\providecommand \href [0]{\begingroup \@sanitize@url \@href}%
\providecommand \@href[1]{\@@startlink{#1}\@@href}%
\providecommand \@@href[1]{\endgroup#1\@@endlink}%
\providecommand \@sanitize@url [0]{\catcode `\\12\catcode `\$12\catcode
  `\&12\catcode `\#12\catcode `\^12\catcode `\_12\catcode `\%12\relax}%
\providecommand \@@startlink[1]{}%
\providecommand \@@endlink[0]{}%
\providecommand \url  [0]{\begingroup\@sanitize@url \@url }%
\providecommand \@url [1]{\endgroup\@href {#1}{\urlprefix }}%
\providecommand \urlprefix  [0]{URL }%
\providecommand \Eprint [0]{\href }%
\providecommand \doibase [0]{http://dx.doi.org/}%
\providecommand \selectlanguage [0]{\@gobble}%
\providecommand \bibinfo  [0]{\@secondoftwo}%
\providecommand \bibfield  [0]{\@secondoftwo}%
\providecommand \translation [1]{[#1]}%
\providecommand \BibitemOpen [0]{}%
\providecommand \bibitemStop [0]{}%
\providecommand \bibitemNoStop [0]{.\EOS\space}%
\providecommand \EOS [0]{\spacefactor3000\relax}%
\providecommand \BibitemShut  [1]{\csname bibitem#1\endcsname}%
\let\auto@bib@innerbib\@empty
\bibitem [{\citenamefont {Drewsen}(2015)}]{Drewsen2015}%
  \BibitemOpen
  \bibfield  {author} {\bibinfo {author} {\bibfnamefont {M.}~\bibnamefont
  {Drewsen}},\ }\href
  {http://phys.au.dk/fileadmin/site_files/forskning/iontrap/pdfs/IonCoulombCrystals_Drewsen_2015.pdf}
  {\bibfield  {journal} {\bibinfo  {journal} {Physica B: Condensed Matter}\
  }\textbf {\bibinfo {volume} {460}},\ \bibinfo {pages} {105} (\bibinfo {year}
  {2015})}\BibitemShut {NoStop}%
\bibitem [{\citenamefont {Paul}(1990)}]{Paul1990}%
  \BibitemOpen
  \bibfield  {author} {\bibinfo {author} {\bibfnamefont {W.}~\bibnamefont
  {Paul}},\ }\href {\doibase 10.1103/revmodphys.62.531} {\bibfield  {journal}
  {\bibinfo  {journal} {Reviews of Modern Physics}\ }\textbf {\bibinfo {volume}
  {62}},\ \bibinfo {pages} {531} (\bibinfo {year} {1990})}\BibitemShut
  {NoStop}%
\bibitem [{\citenamefont {Dehmelt}(1990)}]{Dehmelt1990}%
  \BibitemOpen
  \bibfield  {author} {\bibinfo {author} {\bibfnamefont {H.}~\bibnamefont
  {Dehmelt}},\ }\href {\doibase 10.1103/RevModPhys.62.525} {\bibfield
  {journal} {\bibinfo  {journal} {Rev. Mod. Phys.}\ }\textbf {\bibinfo {volume}
  {62}},\ \bibinfo {pages} {525} (\bibinfo {year} {1990})}\BibitemShut
  {NoStop}%
\bibitem [{\citenamefont {Caillol}\ \emph {et~al.}(1982)\citenamefont
  {Caillol}, \citenamefont {Levesque}, \citenamefont {Weis},\ and\
  \citenamefont {Hansen}}]{Caillol1982}%
  \BibitemOpen
  \bibfield  {author} {\bibinfo {author} {\bibfnamefont {J.~M.}\ \bibnamefont
  {Caillol}}, \bibinfo {author} {\bibfnamefont {D.}~\bibnamefont {Levesque}},
  \bibinfo {author} {\bibfnamefont {J.~J.}\ \bibnamefont {Weis}}, \ and\
  \bibinfo {author} {\bibfnamefont {J.~P.}\ \bibnamefont {Hansen}},\ }\href
  {\doibase 10.1007/BF01012609} {\bibfield  {journal} {\bibinfo  {journal}
  {Journal of Statistical Physics}\ }\textbf {\bibinfo {volume} {28}},\
  \bibinfo {pages} {325} (\bibinfo {year} {1982})}\BibitemShut {NoStop}%
\bibitem [{\citenamefont {Dubin}\ and\ \citenamefont
  {O'Neil}(1999)}]{Dubin1999}%
  \BibitemOpen
  \bibfield  {author} {\bibinfo {author} {\bibfnamefont {D.~H.~E.}\
  \bibnamefont {Dubin}}\ and\ \bibinfo {author} {\bibfnamefont {T.~M.}\
  \bibnamefont {O'Neil}},\ }\href {\doibase 10.1103/RevModPhys.71.87}
  {\bibfield  {journal} {\bibinfo  {journal} {Rev. Mod. Phys.}\ }\textbf
  {\bibinfo {volume} {71}},\ \bibinfo {pages} {87} (\bibinfo {year}
  {1999})}\BibitemShut {NoStop}%
\bibitem [{\citenamefont {Leibfried}\ \emph {et~al.}(2003)\citenamefont
  {Leibfried}, \citenamefont {Blatt}, \citenamefont {Monroe},\ and\
  \citenamefont {Wineland}}]{Leibfried2003}%
  \BibitemOpen
  \bibfield  {author} {\bibinfo {author} {\bibfnamefont {D.}~\bibnamefont
  {Leibfried}}, \bibinfo {author} {\bibfnamefont {R.}~\bibnamefont {Blatt}},
  \bibinfo {author} {\bibfnamefont {C.}~\bibnamefont {Monroe}}, \ and\ \bibinfo
  {author} {\bibfnamefont {D.}~\bibnamefont {Wineland}},\ }\href {\doibase
  10.1103/RevModPhys.75.281} {\bibfield  {journal} {\bibinfo  {journal} {Rev.
  Mod. Phys.}\ }\textbf {\bibinfo {volume} {75}},\ \bibinfo {pages} {281}
  (\bibinfo {year} {2003})}\BibitemShut {NoStop}%
\bibitem [{\citenamefont {Ballance}\ \emph {et~al.}(2016)\citenamefont
  {Ballance}, \citenamefont {Harty}, \citenamefont {Linke}, \citenamefont
  {Sepiol},\ and\ \citenamefont {Lucas}}]{Ballance2016}%
  \BibitemOpen
  \bibfield  {author} {\bibinfo {author} {\bibfnamefont {C.~J.}\ \bibnamefont
  {Ballance}}, \bibinfo {author} {\bibfnamefont {T.~P.}\ \bibnamefont {Harty}},
  \bibinfo {author} {\bibfnamefont {N.~M.}\ \bibnamefont {Linke}}, \bibinfo
  {author} {\bibfnamefont {M.~A.}\ \bibnamefont {Sepiol}}, \ and\ \bibinfo
  {author} {\bibfnamefont {D.~M.}\ \bibnamefont {Lucas}},\ }\href {\doibase
  10.1103/PhysRevLett.117.060504} {\bibfield  {journal} {\bibinfo  {journal}
  {Phys. Rev. Lett.}\ }\textbf {\bibinfo {volume} {117}},\ \bibinfo {pages}
  {060504} (\bibinfo {year} {2016})}\BibitemShut {NoStop}%
\bibitem [{\citenamefont {Chou}\ \emph {et~al.}(2010)\citenamefont {Chou},
  \citenamefont {Hume}, \citenamefont {Koelemeij}, \citenamefont {Wineland},\
  and\ \citenamefont {Rosenband}}]{Chou2010}%
  \BibitemOpen
  \bibfield  {author} {\bibinfo {author} {\bibfnamefont {C.~W.}\ \bibnamefont
  {Chou}}, \bibinfo {author} {\bibfnamefont {D.~B.}\ \bibnamefont {Hume}},
  \bibinfo {author} {\bibfnamefont {J.~C.~J.}\ \bibnamefont {Koelemeij}},
  \bibinfo {author} {\bibfnamefont {D.~J.}\ \bibnamefont {Wineland}}, \ and\
  \bibinfo {author} {\bibfnamefont {T.}~\bibnamefont {Rosenband}},\ }\href
  {\doibase 10.1103/PhysRevLett.104.070802} {\bibfield  {journal} {\bibinfo
  {journal} {Phys. Rev. Lett.}\ }\textbf {\bibinfo {volume} {104}},\ \bibinfo
  {pages} {070802} (\bibinfo {year} {2010})}\BibitemShut {NoStop}%
\bibitem [{\citenamefont {Porras}\ and\ \citenamefont
  {Cirac}(2004)}]{Porras2004}%
  \BibitemOpen
  \bibfield  {author} {\bibinfo {author} {\bibfnamefont {D.}~\bibnamefont
  {Porras}}\ and\ \bibinfo {author} {\bibfnamefont {J.~I.}\ \bibnamefont
  {Cirac}},\ }\href {\doibase 10.1103/PhysRevLett.92.207901} {\bibfield
  {journal} {\bibinfo  {journal} {Phys. Rev. Lett.}\ }\textbf {\bibinfo
  {volume} {92}},\ \bibinfo {pages} {207901} (\bibinfo {year}
  {2004})}\BibitemShut {NoStop}%
\bibitem [{\citenamefont {Friedenauer}\ \emph {et~al.}(2008)\citenamefont
  {Friedenauer}, \citenamefont {Schmitz}, \citenamefont {Gl\"uckert},
  \citenamefont {Porras},\ and\ \citenamefont {Schaetz}}]{Friedenauer2008}%
  \BibitemOpen
  \bibfield  {author} {\bibinfo {author} {\bibfnamefont {A.}~\bibnamefont
  {Friedenauer}}, \bibinfo {author} {\bibfnamefont {H.}~\bibnamefont
  {Schmitz}}, \bibinfo {author} {\bibfnamefont {J.~T.}\ \bibnamefont
  {Gl\"uckert}}, \bibinfo {author} {\bibfnamefont {D.}~\bibnamefont {Porras}},
  \ and\ \bibinfo {author} {\bibfnamefont {T.}~\bibnamefont {Schaetz}},\ }\href
  {http://dx.doi.org/10.1038/nphys1032} {\bibfield  {journal} {\bibinfo
  {journal} {Nat. Phys.}\ }\textbf {\bibinfo {volume} {4}},\ \bibinfo {pages}
  {757} (\bibinfo {year} {2008})}\BibitemShut {NoStop}%
\bibitem [{\citenamefont {Zhang}\ \emph {et~al.}(2017)\citenamefont {Zhang},
  \citenamefont {Pagano}, \citenamefont {Hess}, \citenamefont {Kyprianidis},
  \citenamefont {Becker}, \citenamefont {Kaplan}, \citenamefont {Gorshkov},
  \citenamefont {Gong},\ and\ \citenamefont {Monroe}}]{Zhang2017}%
  \BibitemOpen
  \bibfield  {author} {\bibinfo {author} {\bibfnamefont {J.}~\bibnamefont
  {Zhang}}, \bibinfo {author} {\bibfnamefont {G.}~\bibnamefont {Pagano}},
  \bibinfo {author} {\bibfnamefont {P.~W.}\ \bibnamefont {Hess}}, \bibinfo
  {author} {\bibfnamefont {A.}~\bibnamefont {Kyprianidis}}, \bibinfo {author}
  {\bibfnamefont {P.}~\bibnamefont {Becker}}, \bibinfo {author} {\bibfnamefont
  {H.}~\bibnamefont {Kaplan}}, \bibinfo {author} {\bibfnamefont {A.~V.}\
  \bibnamefont {Gorshkov}}, \bibinfo {author} {\bibfnamefont {Z.-X.}\
  \bibnamefont {Gong}}, \ and\ \bibinfo {author} {\bibfnamefont
  {C.}~\bibnamefont {Monroe}},\ }\href {http://dx.doi.org/10.1038/nature24654}
  {\bibfield  {journal} {\bibinfo  {journal} {Nature}\ }\textbf {\bibinfo
  {volume} {551}},\ \bibinfo {pages} {601 EP } (\bibinfo {year}
  {2017})}\BibitemShut {NoStop}%
\bibitem [{\citenamefont {Berkeland}\ \emph {et~al.}(1998)\citenamefont
  {Berkeland}, \citenamefont {Miller}, \citenamefont {Bergquist}, \citenamefont
  {Itano},\ and\ \citenamefont {Wineland}}]{Berkeland1998}%
  \BibitemOpen
  \bibfield  {author} {\bibinfo {author} {\bibfnamefont {D.}~\bibnamefont
  {Berkeland}}, \bibinfo {author} {\bibfnamefont {J.}~\bibnamefont {Miller}},
  \bibinfo {author} {\bibfnamefont {J.}~\bibnamefont {Bergquist}}, \bibinfo
  {author} {\bibfnamefont {W.}~\bibnamefont {Itano}}, \ and\ \bibinfo {author}
  {\bibfnamefont {D.}~\bibnamefont {Wineland}},\ }\href@noop {} {\bibfield
  {journal} {\bibinfo  {journal} {Journal of Applied Physics}\ }\textbf
  {\bibinfo {volume} {83}},\ \bibinfo {pages} {10} (\bibinfo {year}
  {1998})}\BibitemShut {NoStop}%
\bibitem [{\citenamefont {Thompson}(2015)}]{Thompson2015}%
  \BibitemOpen
  \bibfield  {author} {\bibinfo {author} {\bibfnamefont {R.~C.}\ \bibnamefont
  {Thompson}},\ }\href {\doibase 10.1080/00107514.2014.989715} {\bibfield
  {journal} {\bibinfo  {journal} {Contemporary Physics}\ }\textbf {\bibinfo
  {volume} {56}},\ \bibinfo {pages} {63} (\bibinfo {year} {2015})}\BibitemShut
  {NoStop}%
\bibitem [{\citenamefont {Cetina}\ \emph {et~al.}(2012)\citenamefont {Cetina},
  \citenamefont {Grier},\ and\ \citenamefont {Vuleti\ifmmode~\acute{c}\else
  \'{c}\fi{}}}]{Cetina2012}%
  \BibitemOpen
  \bibfield  {author} {\bibinfo {author} {\bibfnamefont {M.}~\bibnamefont
  {Cetina}}, \bibinfo {author} {\bibfnamefont {A.~T.}\ \bibnamefont {Grier}}, \
  and\ \bibinfo {author} {\bibfnamefont {V.}~\bibnamefont
  {Vuleti\ifmmode~\acute{c}\else \'{c}\fi{}}},\ }\href {\doibase
  10.1103/PhysRevLett.109.253201} {\bibfield  {journal} {\bibinfo  {journal}
  {Phys. Rev. Lett.}\ }\textbf {\bibinfo {volume} {109}},\ \bibinfo {pages}
  {253201} (\bibinfo {year} {2012})}\BibitemShut {NoStop}%
\bibitem [{\citenamefont {Meir}\ \emph {et~al.}(2016)\citenamefont {Meir},
  \citenamefont {Sikorsky}, \citenamefont {Ben-shlomi}, \citenamefont
  {Akerman}, \citenamefont {Dallal},\ and\ \citenamefont {Ozeri}}]{Meir2016}%
  \BibitemOpen
  \bibfield  {author} {\bibinfo {author} {\bibfnamefont {Z.}~\bibnamefont
  {Meir}}, \bibinfo {author} {\bibfnamefont {T.}~\bibnamefont {Sikorsky}},
  \bibinfo {author} {\bibfnamefont {R.}~\bibnamefont {Ben-shlomi}}, \bibinfo
  {author} {\bibfnamefont {N.}~\bibnamefont {Akerman}}, \bibinfo {author}
  {\bibfnamefont {Y.}~\bibnamefont {Dallal}}, \ and\ \bibinfo {author}
  {\bibfnamefont {R.}~\bibnamefont {Ozeri}},\ }\href {\doibase
  10.1103/PhysRevLett.117.243401} {\bibfield  {journal} {\bibinfo  {journal}
  {Phys. Rev. Lett.}\ }\textbf {\bibinfo {volume} {117}},\ \bibinfo {pages}
  {243401} (\bibinfo {year} {2016})}\BibitemShut {NoStop}%
\bibitem [{\citenamefont {Grimm}\ \emph {et~al.}(2000)\citenamefont {Grimm},
  \citenamefont {Weidem\"{u}ller},\ and\ \citenamefont
  {Ovchinnikov}}]{Grimm2000}%
  \BibitemOpen
  \bibfield  {author} {\bibinfo {author} {\bibfnamefont {R.}~\bibnamefont
  {Grimm}}, \bibinfo {author} {\bibfnamefont {M.}~\bibnamefont
  {Weidem\"{u}ller}}, \ and\ \bibinfo {author} {\bibfnamefont {Y.}~\bibnamefont
  {Ovchinnikov}},\ }\href@noop {} {\bibfield  {journal} {\bibinfo  {journal}
  {Adv. At. Mol. Opt. Phys.}\ }\textbf {\bibinfo {volume} {42}},\ \bibinfo
  {pages} {95} (\bibinfo {year} {2000})}\BibitemShut {NoStop}%
\bibitem [{\citenamefont {Lambrecht}\ \emph {et~al.}(2017)\citenamefont
  {Lambrecht}, \citenamefont {Schmidt}, \citenamefont {Weckesser},
  \citenamefont {Debatin}, \citenamefont {Karpa},\ and\ \citenamefont
  {Schaetz}}]{Lambrecht2017}%
  \BibitemOpen
  \bibfield  {author} {\bibinfo {author} {\bibfnamefont {A.}~\bibnamefont
  {Lambrecht}}, \bibinfo {author} {\bibfnamefont {J.}~\bibnamefont {Schmidt}},
  \bibinfo {author} {\bibfnamefont {P.}~\bibnamefont {Weckesser}}, \bibinfo
  {author} {\bibfnamefont {M.}~\bibnamefont {Debatin}}, \bibinfo {author}
  {\bibfnamefont {L.}~\bibnamefont {Karpa}}, \ and\ \bibinfo {author}
  {\bibfnamefont {T.}~\bibnamefont {Schaetz}},\ }\href {\doibase
  10.1038/s41566-017-0030-2} {\bibfield  {journal} {\bibinfo  {journal} {Nature
  Photonics}\ }\textbf {\bibinfo {volume} {11}},\ \bibinfo {pages} {704}
  (\bibinfo {year} {2017})}\BibitemShut {NoStop}%
\bibitem [{\citenamefont {Schaetz}(2017)}]{Schaetz2017}%
  \BibitemOpen
  \bibfield  {author} {\bibinfo {author} {\bibfnamefont {T.}~\bibnamefont
  {Schaetz}},\ }\href {http://stacks.iop.org/0953-4075/50/i=10/a=102001}
  {\bibfield  {journal} {\bibinfo  {journal} {Journal of Physics B: Atomic,
  Molecular and Optical Physics}\ }\textbf {\bibinfo {volume} {50}},\ \bibinfo
  {pages} {102001} (\bibinfo {year} {2017})}\BibitemShut {NoStop}%
\bibitem [{\citenamefont {Endres}\ \emph {et~al.}(2016)\citenamefont {Endres},
  \citenamefont {Bernien}, \citenamefont {Keesling}, \citenamefont {Levine},
  \citenamefont {Anschuetz}, \citenamefont {Krajenbrink}, \citenamefont
  {Senko}, \citenamefont {Vuletic}, \citenamefont {Greiner},\ and\
  \citenamefont {Lukin}}]{Endres2016}%
  \BibitemOpen
  \bibfield  {author} {\bibinfo {author} {\bibfnamefont {M.}~\bibnamefont
  {Endres}}, \bibinfo {author} {\bibfnamefont {H.}~\bibnamefont {Bernien}},
  \bibinfo {author} {\bibfnamefont {A.}~\bibnamefont {Keesling}}, \bibinfo
  {author} {\bibfnamefont {H.}~\bibnamefont {Levine}}, \bibinfo {author}
  {\bibfnamefont {E.~R.}\ \bibnamefont {Anschuetz}}, \bibinfo {author}
  {\bibfnamefont {A.}~\bibnamefont {Krajenbrink}}, \bibinfo {author}
  {\bibfnamefont {C.}~\bibnamefont {Senko}}, \bibinfo {author} {\bibfnamefont
  {V.}~\bibnamefont {Vuletic}}, \bibinfo {author} {\bibfnamefont
  {M.}~\bibnamefont {Greiner}}, \ and\ \bibinfo {author} {\bibfnamefont
  {M.~D.}\ \bibnamefont {Lukin}},\ }\href {\doibase 10.1126/science.aah3752}
  {\bibfield  {journal} {\bibinfo  {journal} {Science}\ }\textbf {\bibinfo
  {volume} {354}},\ \bibinfo {pages} {1024} (\bibinfo {year}
  {2016})}\BibitemShut {NoStop}%
\bibitem [{\citenamefont {Cormick}\ \emph {et~al.}(2011)\citenamefont
  {Cormick}, \citenamefont {Schaetz},\ and\ \citenamefont
  {Morigi}}]{Cormick2011}%
  \BibitemOpen
  \bibfield  {author} {\bibinfo {author} {\bibfnamefont {C.}~\bibnamefont
  {Cormick}}, \bibinfo {author} {\bibfnamefont {T.}~\bibnamefont {Schaetz}}, \
  and\ \bibinfo {author} {\bibfnamefont {G.}~\bibnamefont {Morigi}},\ }\href
  {http://stacks.iop.org/1367-2630/13/i=4/a=043019} {\bibfield  {journal}
  {\bibinfo  {journal} {New Journal of Physics}\ }\textbf {\bibinfo {volume}
  {13}},\ \bibinfo {pages} {043019} (\bibinfo {year} {2011})}\BibitemShut
  {NoStop}%
\bibitem [{\citenamefont {Laupr\^{e}tre}\ \emph {et~al.}(2017)\citenamefont
  {Laupr\^{e}tre}, \citenamefont {Linnet}, \citenamefont {D.~Leroux},
  \citenamefont {Dantan},\ and\ \citenamefont {Drewsen}}]{Laupretre2017}%
  \BibitemOpen
  \bibfield  {author} {\bibinfo {author} {\bibfnamefont {T.}~\bibnamefont
  {Laupr\^{e}tre}}, \bibinfo {author} {\bibfnamefont {R.~B.}\ \bibnamefont
  {Linnet}}, \bibinfo {author} {\bibfnamefont {I.}~\bibnamefont {D.~Leroux}},
  \bibinfo {author} {\bibfnamefont {A.}~\bibnamefont {Dantan}}, \ and\ \bibinfo
  {author} {\bibfnamefont {M.}~\bibnamefont {Drewsen}},\ }\href@noop {}
  {\bibfield  {journal} {\bibinfo  {journal} {arXiv preprint
  arXiv:1703.050892}\ } (\bibinfo {year} {2017})}\BibitemShut {NoStop}%
\bibitem [{\citenamefont {Linnet}\ \emph {et~al.}(2012)\citenamefont {Linnet},
  \citenamefont {Leroux}, \citenamefont {Marciante}, \citenamefont {Dantan},\
  and\ \citenamefont {Drewsen}}]{Linnet2012}%
  \BibitemOpen
  \bibfield  {author} {\bibinfo {author} {\bibfnamefont {R.~B.}\ \bibnamefont
  {Linnet}}, \bibinfo {author} {\bibfnamefont {I.~D.}\ \bibnamefont {Leroux}},
  \bibinfo {author} {\bibfnamefont {M.}~\bibnamefont {Marciante}}, \bibinfo
  {author} {\bibfnamefont {A.}~\bibnamefont {Dantan}}, \ and\ \bibinfo {author}
  {\bibfnamefont {M.}~\bibnamefont {Drewsen}},\ }\href {\doibase
  10.1103/PhysRevLett.109.233005} {\bibfield  {journal} {\bibinfo  {journal}
  {Phys. Rev. Lett.}\ }\textbf {\bibinfo {volume} {109}},\ \bibinfo {pages}
  {233005} (\bibinfo {year} {2012})}\BibitemShut {NoStop}%
\bibitem [{\citenamefont {Karpa}\ \emph {et~al.}(2013)\citenamefont {Karpa},
  \citenamefont {Bylinskii}, \citenamefont {Gangloff}, \citenamefont {Cetina},\
  and\ \citenamefont {Vuleti\ifmmode~\acute{c}\else \'{c}\fi{}}}]{Karpa2013}%
  \BibitemOpen
  \bibfield  {author} {\bibinfo {author} {\bibfnamefont {L.}~\bibnamefont
  {Karpa}}, \bibinfo {author} {\bibfnamefont {A.}~\bibnamefont {Bylinskii}},
  \bibinfo {author} {\bibfnamefont {D.}~\bibnamefont {Gangloff}}, \bibinfo
  {author} {\bibfnamefont {M.}~\bibnamefont {Cetina}}, \ and\ \bibinfo {author}
  {\bibfnamefont {V.}~\bibnamefont {Vuleti\ifmmode~\acute{c}\else
  \'{c}\fi{}}},\ }\href {\doibase 10.1103/PhysRevLett.111.163002} {\bibfield
  {journal} {\bibinfo  {journal} {Phys. Rev. Lett.}\ }\textbf {\bibinfo
  {volume} {111}},\ \bibinfo {pages} {163002} (\bibinfo {year}
  {2013})}\BibitemShut {NoStop}%
\bibitem [{\citenamefont {Bylinskii}\ \emph {et~al.}(2015)\citenamefont
  {Bylinskii}, \citenamefont {Gangloff},\ and\ \citenamefont
  {Vuleti{\'c}}}]{Bylinskii2015}%
  \BibitemOpen
  \bibfield  {author} {\bibinfo {author} {\bibfnamefont {A.}~\bibnamefont
  {Bylinskii}}, \bibinfo {author} {\bibfnamefont {D.}~\bibnamefont {Gangloff}},
  \ and\ \bibinfo {author} {\bibfnamefont {V.}~\bibnamefont {Vuleti{\'c}}},\
  }\href {\doibase 10.1126/science.1261422} {\bibfield  {journal} {\bibinfo
  {journal} {Science}\ }\textbf {\bibinfo {volume} {348}},\ \bibinfo {pages}
  {1115} (\bibinfo {year} {2015})}\BibitemShut {NoStop}%
\bibitem [{\citenamefont {Leschhorn}\ \emph {et~al.}(2012)\citenamefont
  {Leschhorn}, \citenamefont {Hasegawa},\ and\ \citenamefont
  {Schaetz}}]{Leschhorn2012}%
  \BibitemOpen
  \bibfield  {author} {\bibinfo {author} {\bibfnamefont {G.}~\bibnamefont
  {Leschhorn}}, \bibinfo {author} {\bibfnamefont {T.}~\bibnamefont {Hasegawa}},
  \ and\ \bibinfo {author} {\bibfnamefont {T.}~\bibnamefont {Schaetz}},\
  }\href@noop {} {\bibfield  {journal} {\bibinfo  {journal} {Appl. Phys. B}\
  }\textbf {\bibinfo {volume} {108}},\ \bibinfo {pages} {159} (\bibinfo {year}
  {2012})}\BibitemShut {NoStop}%
\bibitem [{\citenamefont {Huber}\ \emph {et~al.}(2014)\citenamefont {Huber},
  \citenamefont {Lambrecht}, \citenamefont {Schmidt}, \citenamefont {Karpa},\
  and\ \citenamefont {Schaetz}}]{Huber2014}%
  \BibitemOpen
  \bibfield  {author} {\bibinfo {author} {\bibfnamefont {T.}~\bibnamefont
  {Huber}}, \bibinfo {author} {\bibfnamefont {A.}~\bibnamefont {Lambrecht}},
  \bibinfo {author} {\bibfnamefont {J.}~\bibnamefont {Schmidt}}, \bibinfo
  {author} {\bibfnamefont {L.}~\bibnamefont {Karpa}}, \ and\ \bibinfo {author}
  {\bibfnamefont {T.}~\bibnamefont {Schaetz}},\ }\href@noop {} {\bibfield
  {journal} {\bibinfo  {journal} {Nature Communications}\ }\textbf {\bibinfo
  {volume} {5}} (\bibinfo {year} {2014})}\BibitemShut {NoStop}%
\bibitem [{\citenamefont {Schneider}\ \emph {et~al.}(2010)\citenamefont
  {Schneider}, \citenamefont {Enderlein}, \citenamefont {Huber},\ and\
  \citenamefont {Schaetz}}]{Schneider2010}%
  \BibitemOpen
  \bibfield  {author} {\bibinfo {author} {\bibfnamefont {C.}~\bibnamefont
  {Schneider}}, \bibinfo {author} {\bibfnamefont {M.}~\bibnamefont
  {Enderlein}}, \bibinfo {author} {\bibfnamefont {T.}~\bibnamefont {Huber}}, \
  and\ \bibinfo {author} {\bibfnamefont {T.}~\bibnamefont {Schaetz}},\ }\href
  {http://dx.doi.org/10.1038/nphoton.2010.236} {\bibfield  {journal} {\bibinfo
  {journal} {Nature Photonics}\ }\textbf {\bibinfo {volume} {4}},\ \bibinfo
  {pages} {772} (\bibinfo {year} {2010})}\BibitemShut {NoStop}%
\bibitem [{\citenamefont {Wilson}(1927)}]{Wilson1927}%
  \BibitemOpen
  \bibfield  {author} {\bibinfo {author} {\bibfnamefont {E.~B.}\ \bibnamefont
  {Wilson}},\ }\href {\doibase 10.1080/01621459.1927.10502953} {\bibfield
  {journal} {\bibinfo  {journal} {Journal of the American Statistical
  Association}\ }\textbf {\bibinfo {volume} {22}},\ \bibinfo {pages} {209}
  (\bibinfo {year} {1927})}\BibitemShut {NoStop}%
\bibitem [{\citenamefont {Schneider}\ \emph
  {et~al.}(2012{\natexlab{a}})\citenamefont {Schneider}, \citenamefont
  {Enderlein}, \citenamefont {Huber}, \citenamefont {D\"urr},\ and\
  \citenamefont {Schaetz}}]{Schneider2012}%
  \BibitemOpen
  \bibfield  {author} {\bibinfo {author} {\bibfnamefont {C.}~\bibnamefont
  {Schneider}}, \bibinfo {author} {\bibfnamefont {M.}~\bibnamefont
  {Enderlein}}, \bibinfo {author} {\bibfnamefont {T.}~\bibnamefont {Huber}},
  \bibinfo {author} {\bibfnamefont {S.}~\bibnamefont {D\"urr}}, \ and\ \bibinfo
  {author} {\bibfnamefont {T.}~\bibnamefont {Schaetz}},\ }\href {\doibase
  10.1103/PhysRevA.85.013422} {\bibfield  {journal} {\bibinfo  {journal} {Phys.
  Rev. A}\ }\textbf {\bibinfo {volume} {85}},\ \bibinfo {pages} {013422}
  (\bibinfo {year} {2012}{\natexlab{a}})}\BibitemShut {NoStop}%
\bibitem [{\citenamefont {Lindemann}(1910)}]{Lindemann1910}%
  \BibitemOpen
  \bibfield  {author} {\bibinfo {author} {\bibfnamefont {F.~A.}\ \bibnamefont
  {Lindemann}},\ }\href@noop {} {\bibfield  {journal} {\bibinfo  {journal}
  {Phys. Z.,(West Germany)}\ }\textbf {\bibinfo {volume} {11}},\ \bibinfo
  {pages} {609} (\bibinfo {year} {1910})}\BibitemShut {NoStop}%
\bibitem [{\citenamefont {Wineland}(2013)}]{Wineland2013}%
  \BibitemOpen
  \bibfield  {author} {\bibinfo {author} {\bibfnamefont {D.~J.}\ \bibnamefont
  {Wineland}},\ }\href {\doibase 10.1103/revmodphys.85.1103} {\bibfield
  {journal} {\bibinfo  {journal} {Reviews of Modern Physics}\ }\textbf
  {\bibinfo {volume} {85}},\ \bibinfo {pages} {1103} (\bibinfo {year}
  {2013})}\BibitemShut {NoStop}%
\bibitem [{\citenamefont {Clos}\ \emph {et~al.}(2016)\citenamefont {Clos},
  \citenamefont {Porras}, \citenamefont {Warring},\ and\ \citenamefont
  {Schaetz}}]{Clos2016}%
  \BibitemOpen
  \bibfield  {author} {\bibinfo {author} {\bibfnamefont {G.}~\bibnamefont
  {Clos}}, \bibinfo {author} {\bibfnamefont {D.}~\bibnamefont {Porras}},
  \bibinfo {author} {\bibfnamefont {U.}~\bibnamefont {Warring}}, \ and\
  \bibinfo {author} {\bibfnamefont {T.}~\bibnamefont {Schaetz}},\ }\href
  {\doibase 10.1103/PhysRevLett.117.170401} {\bibfield  {journal} {\bibinfo
  {journal} {Phys. Rev. Lett.}\ }\textbf {\bibinfo {volume} {117}},\ \bibinfo
  {pages} {170401} (\bibinfo {year} {2016})}\BibitemShut {NoStop}%
\bibitem [{\citenamefont {Schneider}\ \emph
  {et~al.}(2012{\natexlab{b}})\citenamefont {Schneider}, \citenamefont
  {Porras},\ and\ \citenamefont {Schaetz}}]{Schneider2012b}%
  \BibitemOpen
  \bibfield  {author} {\bibinfo {author} {\bibfnamefont {C.}~\bibnamefont
  {Schneider}}, \bibinfo {author} {\bibfnamefont {D.}~\bibnamefont {Porras}}, \
  and\ \bibinfo {author} {\bibfnamefont {T.}~\bibnamefont {Schaetz}},\ }\href
  {http://stacks.iop.org/0034-4885/75/i=2/a=024401} {\bibfield  {journal}
  {\bibinfo  {journal} {Reports on Progress in Physics}\ }\textbf {\bibinfo
  {volume} {75}},\ \bibinfo {pages} {024401} (\bibinfo {year}
  {2012}{\natexlab{b}})}\BibitemShut {NoStop}%
\bibitem [{\citenamefont {Bermudez}\ \emph {et~al.}(2012)\citenamefont
  {Bermudez}, \citenamefont {Schaetz},\ and\ \citenamefont
  {Porras}}]{Bermudez2012}%
  \BibitemOpen
  \bibfield  {author} {\bibinfo {author} {\bibfnamefont {A.}~\bibnamefont
  {Bermudez}}, \bibinfo {author} {\bibfnamefont {T.}~\bibnamefont {Schaetz}}, \
  and\ \bibinfo {author} {\bibfnamefont {D.}~\bibnamefont {Porras}},\ }\href
  {http://stacks.iop.org/1367-2630/14/i=5/a=053049} {\bibfield  {journal}
  {\bibinfo  {journal} {New Journal of Physics}\ }\textbf {\bibinfo {volume}
  {14}},\ \bibinfo {pages} {053049} (\bibinfo {year} {2012})}\BibitemShut
  {NoStop}%
\bibitem [{\citenamefont {Retzker}\ \emph {et~al.}(2008)\citenamefont
  {Retzker}, \citenamefont {Thompson}, \citenamefont {Segal},\ and\
  \citenamefont {Plenio}}]{Retzker2008}%
  \BibitemOpen
  \bibfield  {author} {\bibinfo {author} {\bibfnamefont {A.}~\bibnamefont
  {Retzker}}, \bibinfo {author} {\bibfnamefont {R.~C.}\ \bibnamefont
  {Thompson}}, \bibinfo {author} {\bibfnamefont {D.~M.}\ \bibnamefont {Segal}},
  \ and\ \bibinfo {author} {\bibfnamefont {M.~B.}\ \bibnamefont {Plenio}},\
  }\href {\doibase 10.1103/PhysRevLett.101.260504} {\bibfield  {journal}
  {\bibinfo  {journal} {Phys. Rev. Lett.}\ }\textbf {\bibinfo {volume} {101}},\
  \bibinfo {pages} {260504} (\bibinfo {year} {2008})}\BibitemShut {NoStop}%
\bibitem [{\citenamefont {Shimshoni}\ \emph
  {et~al.}(2011{\natexlab{a}})\citenamefont {Shimshoni}, \citenamefont
  {Morigi},\ and\ \citenamefont {Fishman}}]{Shimshoni2011}%
  \BibitemOpen
  \bibfield  {author} {\bibinfo {author} {\bibfnamefont {E.}~\bibnamefont
  {Shimshoni}}, \bibinfo {author} {\bibfnamefont {G.}~\bibnamefont {Morigi}}, \
  and\ \bibinfo {author} {\bibfnamefont {S.}~\bibnamefont {Fishman}},\ }\href
  {\doibase 10.1103/PhysRevLett.106.010401} {\bibfield  {journal} {\bibinfo
  {journal} {Phys. Rev. Lett.}\ }\textbf {\bibinfo {volume} {106}},\ \bibinfo
  {pages} {010401} (\bibinfo {year} {2011}{\natexlab{a}})}\BibitemShut
  {NoStop}%
\bibitem [{\citenamefont {Shimshoni}\ \emph
  {et~al.}(2011{\natexlab{b}})\citenamefont {Shimshoni}, \citenamefont
  {Morigi},\ and\ \citenamefont {Fishman}}]{Simshoni2011b}%
  \BibitemOpen
  \bibfield  {author} {\bibinfo {author} {\bibfnamefont {E.}~\bibnamefont
  {Shimshoni}}, \bibinfo {author} {\bibfnamefont {G.}~\bibnamefont {Morigi}}, \
  and\ \bibinfo {author} {\bibfnamefont {S.}~\bibnamefont {Fishman}},\ }\href
  {\doibase 10.1103/PhysRevA.83.032308} {\bibfield  {journal} {\bibinfo
  {journal} {Phys. Rev. A}\ }\textbf {\bibinfo {volume} {83}},\ \bibinfo
  {pages} {032308} (\bibinfo {year} {2011}{\natexlab{b}})}\BibitemShut
  {NoStop}%
\bibitem [{\citenamefont {Baltrusch}\ \emph {et~al.}(2011)\citenamefont
  {Baltrusch}, \citenamefont {Cormick}, \citenamefont {De~Chiara},
  \citenamefont {Calarco},\ and\ \citenamefont {Morigi}}]{Baltrusch2011}%
  \BibitemOpen
  \bibfield  {author} {\bibinfo {author} {\bibfnamefont {J.~D.}\ \bibnamefont
  {Baltrusch}}, \bibinfo {author} {\bibfnamefont {C.}~\bibnamefont {Cormick}},
  \bibinfo {author} {\bibfnamefont {G.}~\bibnamefont {De~Chiara}}, \bibinfo
  {author} {\bibfnamefont {T.}~\bibnamefont {Calarco}}, \ and\ \bibinfo
  {author} {\bibfnamefont {G.}~\bibnamefont {Morigi}},\ }\href {\doibase
  10.1103/PhysRevA.84.063821} {\bibfield  {journal} {\bibinfo  {journal} {Phys.
  Rev. A}\ }\textbf {\bibinfo {volume} {84}},\ \bibinfo {pages} {063821}
  (\bibinfo {year} {2011})}\BibitemShut {NoStop}%
\bibitem [{\citenamefont {Baltrusch}\ \emph {et~al.}(2012)\citenamefont
  {Baltrusch}, \citenamefont {Cormick},\ and\ \citenamefont
  {Morigi}}]{Baltrusch2012}%
  \BibitemOpen
  \bibfield  {author} {\bibinfo {author} {\bibfnamefont {J.~D.}\ \bibnamefont
  {Baltrusch}}, \bibinfo {author} {\bibfnamefont {C.}~\bibnamefont {Cormick}},
  \ and\ \bibinfo {author} {\bibfnamefont {G.}~\bibnamefont {Morigi}},\ }\href
  {\doibase 10.1103/PhysRevA.86.032104} {\bibfield  {journal} {\bibinfo
  {journal} {Phys. Rev. A}\ }\textbf {\bibinfo {volume} {86}},\ \bibinfo
  {pages} {032104} (\bibinfo {year} {2012})}\BibitemShut {NoStop}%
\bibitem [{\citenamefont {C\^ot\'e}\ \emph {et~al.}(2002)\citenamefont
  {C\^ot\'e}, \citenamefont {Kharchenko},\ and\ \citenamefont
  {Lukin}}]{Cote2002}%
  \BibitemOpen
  \bibfield  {author} {\bibinfo {author} {\bibfnamefont {R.}~\bibnamefont
  {C\^ot\'e}}, \bibinfo {author} {\bibfnamefont {V.}~\bibnamefont
  {Kharchenko}}, \ and\ \bibinfo {author} {\bibfnamefont {M.~D.}\ \bibnamefont
  {Lukin}},\ }\href {\doibase 10.1103/PhysRevLett.89.093001} {\bibfield
  {journal} {\bibinfo  {journal} {Phys. Rev. Lett.}\ }\textbf {\bibinfo
  {volume} {89}},\ \bibinfo {pages} {093001} (\bibinfo {year}
  {2002})}\BibitemShut {NoStop}%
\bibitem [{\citenamefont {Schurer}\ \emph {et~al.}(2017)\citenamefont
  {Schurer}, \citenamefont {Negretti},\ and\ \citenamefont
  {Schmelcher}}]{Schurer2017}%
  \BibitemOpen
  \bibfield  {author} {\bibinfo {author} {\bibfnamefont {J.~M.}\ \bibnamefont
  {Schurer}}, \bibinfo {author} {\bibfnamefont {A.}~\bibnamefont {Negretti}}, \
  and\ \bibinfo {author} {\bibfnamefont {P.}~\bibnamefont {Schmelcher}},\
  }\href {\doibase 10.1103/PhysRevLett.119.063001} {\bibfield  {journal}
  {\bibinfo  {journal} {Phys. Rev. Lett.}\ }\textbf {\bibinfo {volume} {119}},\
  \bibinfo {pages} {063001} (\bibinfo {year} {2017})}\BibitemShut {NoStop}%
\bibitem [{\citenamefont {Tomza}\ \emph {et~al.}(2017)\citenamefont {Tomza},
  \citenamefont {Jachymski}, \citenamefont {Gerritsma}, \citenamefont
  {Negretti}, \citenamefont {Calarco}, \citenamefont {Idziaszek},\ and\
  \citenamefont {Julienne}}]{Tomza2017}%
  \BibitemOpen
  \bibfield  {author} {\bibinfo {author} {\bibfnamefont {M.}~\bibnamefont
  {Tomza}}, \bibinfo {author} {\bibfnamefont {K.}~\bibnamefont {Jachymski}},
  \bibinfo {author} {\bibfnamefont {R.}~\bibnamefont {Gerritsma}}, \bibinfo
  {author} {\bibfnamefont {A.}~\bibnamefont {Negretti}}, \bibinfo {author}
  {\bibfnamefont {T.}~\bibnamefont {Calarco}}, \bibinfo {author} {\bibfnamefont
  {Z.}~\bibnamefont {Idziaszek}}, \ and\ \bibinfo {author} {\bibfnamefont
  {P.~S.}\ \bibnamefont {Julienne}},\ }\href@noop {} {\bibfield  {journal}
  {\bibinfo  {journal} {arXiv preprint arXiv:1708.07832}\ } (\bibinfo {year}
  {2017})}\BibitemShut {NoStop}%
\bibitem [{\citenamefont {Bissbort}\ \emph {et~al.}(2013)\citenamefont
  {Bissbort}, \citenamefont {Cocks}, \citenamefont {Negretti}, \citenamefont
  {Idziaszek}, \citenamefont {Calarco}, \citenamefont {Schmidt-Kaler},
  \citenamefont {Hofstetter},\ and\ \citenamefont {Gerritsma}}]{Bissbort2013}%
  \BibitemOpen
  \bibfield  {author} {\bibinfo {author} {\bibfnamefont {U.}~\bibnamefont
  {Bissbort}}, \bibinfo {author} {\bibfnamefont {D.}~\bibnamefont {Cocks}},
  \bibinfo {author} {\bibfnamefont {A.}~\bibnamefont {Negretti}}, \bibinfo
  {author} {\bibfnamefont {Z.}~\bibnamefont {Idziaszek}}, \bibinfo {author}
  {\bibfnamefont {T.}~\bibnamefont {Calarco}}, \bibinfo {author} {\bibfnamefont
  {F.}~\bibnamefont {Schmidt-Kaler}}, \bibinfo {author} {\bibfnamefont
  {W.}~\bibnamefont {Hofstetter}}, \ and\ \bibinfo {author} {\bibfnamefont
  {R.}~\bibnamefont {Gerritsma}},\ }\href {\doibase
  10.1103/PhysRevLett.111.080501} {\bibfield  {journal} {\bibinfo  {journal}
  {Phys. Rev. Lett.}\ }\textbf {\bibinfo {volume} {111}},\ \bibinfo {pages}
  {080501} (\bibinfo {year} {2013})}\BibitemShut {NoStop}%
\bibitem [{\citenamefont {Gerritsma}\ \emph {et~al.}(2012)\citenamefont
  {Gerritsma}, \citenamefont {Negretti}, \citenamefont {Doerk}, \citenamefont
  {Idziaszek}, \citenamefont {Calarco},\ and\ \citenamefont
  {Schmidt-Kaler}}]{Gerritsma2012}%
  \BibitemOpen
  \bibfield  {author} {\bibinfo {author} {\bibfnamefont {R.}~\bibnamefont
  {Gerritsma}}, \bibinfo {author} {\bibfnamefont {A.}~\bibnamefont {Negretti}},
  \bibinfo {author} {\bibfnamefont {H.}~\bibnamefont {Doerk}}, \bibinfo
  {author} {\bibfnamefont {Z.}~\bibnamefont {Idziaszek}}, \bibinfo {author}
  {\bibfnamefont {T.}~\bibnamefont {Calarco}}, \ and\ \bibinfo {author}
  {\bibfnamefont {F.}~\bibnamefont {Schmidt-Kaler}},\ }\href {\doibase
  10.1103/PhysRevLett.109.080402} {\bibfield  {journal} {\bibinfo  {journal}
  {Phys. Rev. Lett.}\ }\textbf {\bibinfo {volume} {109}},\ \bibinfo {pages}
  {080402} (\bibinfo {year} {2012})}\BibitemShut {NoStop}%
\bibitem [{\citenamefont {Ostermann}\ \emph {et~al.}(2016)\citenamefont
  {Ostermann}, \citenamefont {Piazza},\ and\ \citenamefont
  {Ritsch}}]{Ostermann2016}%
  \BibitemOpen
  \bibfield  {author} {\bibinfo {author} {\bibfnamefont {S.}~\bibnamefont
  {Ostermann}}, \bibinfo {author} {\bibfnamefont {F.}~\bibnamefont {Piazza}}, \
  and\ \bibinfo {author} {\bibfnamefont {H.}~\bibnamefont {Ritsch}},\ }\href
  {\doibase 10.1103/PhysRevX.6.021026} {\bibfield  {journal} {\bibinfo
  {journal} {Phys. Rev. X}\ }\textbf {\bibinfo {volume} {6}},\ \bibinfo {pages}
  {021026} (\bibinfo {year} {2016})}\BibitemShut {NoStop}%
\end{thebibliography}
\end{document}